\documentclass[twoside,slac_one]{revtex4}
\usepackage{graphicx}
\usepackage{fancyhdr}
\usepackage{amsmath} % American Mathematics Society standards
\usepackage{bm}% bold math
\usepackage{amsxtra}
\usepackage{amssymb}
\usepackage{amsthm}
\usepackage{latexsym}
\usepackage{lscape}

\begin{document}

\title{ Review of Heavy Flavor Physics at the Tevatron }

\author{Gavril Giurgiu (on behalf of the D0 and CDF Collaborations)}
\affiliation{Department of Physics and Astronomy, Johns Hopkins University, Baltimore, MD, USA}

\begin{abstract}

The D0 and CDF II detectors at the Fermilab Tevatron have each accumulated more 
that 9~fb$^{-1}$ of integrated luminosity. The corresponding large datasets 
enable the two experiments to perform unprecedented studies of heavy flavor 
hadron properties. We present recent D0 and CDF measurements, focusing on rare 
decays and CP violation in $B$-meson decays. 

\end{abstract}

\maketitle

\thispagestyle{fancy}

\section{Introduction}

Flavor Physics probes new phenomena by either searching for small 
deviations from the Standard Model (SM) based theoretical predictions 
or by measuring quantities which are highly suppressed within the SM. 
Searches for small deviations from the SM are performed using large 
strange, charm or bottom hadron samples, mostly by kaon experiments of $B$ factories. 
Measurements of highly suppressed quantities, such as CP violation phases 
and asymmetries in the neutral $B_s$-meson system or searches for rare 
$B$ decays, are performed with the hope that new physics effects would 
be large enough to significantly affect the measured quantities and so  
lead to observations of deviations from the SM expectations. 
    
The D0 and CDF detectors at the Fermilab Tevatron have each accumulated more 
that 9~fb$^{-1}$ of integrated luminosity. The corresponding large datasets 
enable the two experiments to perform unprecedented studies of heavy flavor 
hadron properties. We present recent D0 and CDF measurements, focusing on
rare decays and CP violation in $B$-meson decays. 

\section{Search for $B^0_s \rightarrow \mu^{+} \mu^{-}$ Decays}

Processes involving Flavor Changing Neutral Currents (FCNC) provide excellent opportunities 
to search for evidence of new physics since in the SM they are forbidden at tree level and can 
only occur through higher order loop diagrams. Two such processes are the decays 
$B_{s/d} \rightarrow \mu^{+} \mu^{-}$. 
In the SM, these decays are both Cabibbo and helicity suppressed. Their branching ratios 
are predicted with~10\% accuracy~\cite{buras} as:
$BR(B_{s} \rightarrow \mu^{+} \mu^{-}) = (3.2 \pm 0.2) \times 10^{-9}$ and 
$BR(B_{d} \rightarrow \mu^{+} \mu^{-}) = (1.0 \pm 0.1) \times 10^{-10}$.

These predictions are one order of magnitude smaller than the current experimental 
sensitivity. Enhancements to the expected $B_s \rightarrow \mu^{+} \mu^{-}$ branching fraction 
occur in a variety of different new physics models. For example, in supersymmetry (SUSY) models, 
new supersymmetric particles can increase the branching fraction $BR(B_{s} \rightarrow \mu^{+} \mu^{-})$ by several orders 
of magnitude at large tan$(\beta)$, the ratio of vacuum expectation values of the Higgs doublets~\cite{choudhury}. 
In the minimal supersymmetric standard model (MSSM), the enhancement is proportional to tan$^6(\beta)$. 
For large tan$(\beta)$, this search is one of the most sensitive probes of new physics available 
at the Tevatron experiments.

Using 6.1~fb$^{-1}$, the D0 experiment has published~\cite{D0_Bs_mumu} in 2010 an upper limit on the 
$B_{s} \rightarrow \mu^{+} \mu^{-}$ branching ratio of $51 \times 10^{-9}$ at 95\%CL. 
The CDF experiment has performed a recent update~\cite{CDF_Bs_mumu} of the analysis 
using 7~fb$^{-1}$ of integrated luminosity which supersedes the previous CDF published 
result~\cite{CDF_Bs_mumu_2fb} which used 2~fb$^{-1}$ of data.

In addition to increasing the size of the data set, the sensitivity of this analysis is improved 
another 20\% by including events which cross regions of the tracker where the trigger efficiency is 
rapidly changing and by including events with muons in the forward regions. Other improvements include 
the use of a better neural network (NN) discriminant that provides approximately twice the background 
rejection for the same signal efficiency.

The events are collected using a set of dimuon triggers and must satisfy either of two sets of 
requirements corresponding to different topologies: CC events have both muon candidates detected in the 
central region (CMU), while CF events have one central muon and another muon 
detected in the forward region (CMX).

The baseline selection requires high quality muon candidates with transverse momentum relative to the 
beam direction of $p_T > 2.0 (2.2)$~GeV/c in the central (forward) region. The muon pairs are required to 
have an invariant mass in the range $4.669 < $M$(\mu\mu) < 5.969$~GeV/c$^2$ and are constrained to originate from 
a common well-measured three-dimensional (3D) vertex. A likelihood method together with an energy loss based 
selection are used to further suppress contributions from hadrons misidentified as muons. A fraction 
of the total number of background and simulated signal events are used to train a NN 
to discriminate signal from background events. The remainder are 
used to test for NN over-training and to determine the signal and background efficiencies.

To exploit the difference in the M$(\mu\mu)$ distributions between 
signal and background and the improved suppression of combinatorial background at large NN output ($\nu_{NN}$), the data 
is divided into sub-samples in the ($\nu_{NN}$, M$(\mu\mu)$) plane. The CC and CF samples are each divided 
into 40 sub-samples. There are eight bins in the $\nu_{NN}$. Within each $\nu_{NN}$ bin,  
five M$(\mu\mu)$ bins are employed, each 24 MeV/c$^2$ wide, centered on the world average $B_s$ ($B_d$) mass.

The number of observed events is compared to the number expected in all 80 sub-samples for 
the $B_d$ search region. The data are consistent with the background expectations and yield 
an observed limit of 
\begin{center}
$BR(B_d \rightarrow \mu^+\mu^-)~<~6.0~(5.0)~\times~10^{-9}$ at 95\% (90\%) C.L. 
\end{center}

\begin{figure}[tb]
\includegraphics[width=80mm]{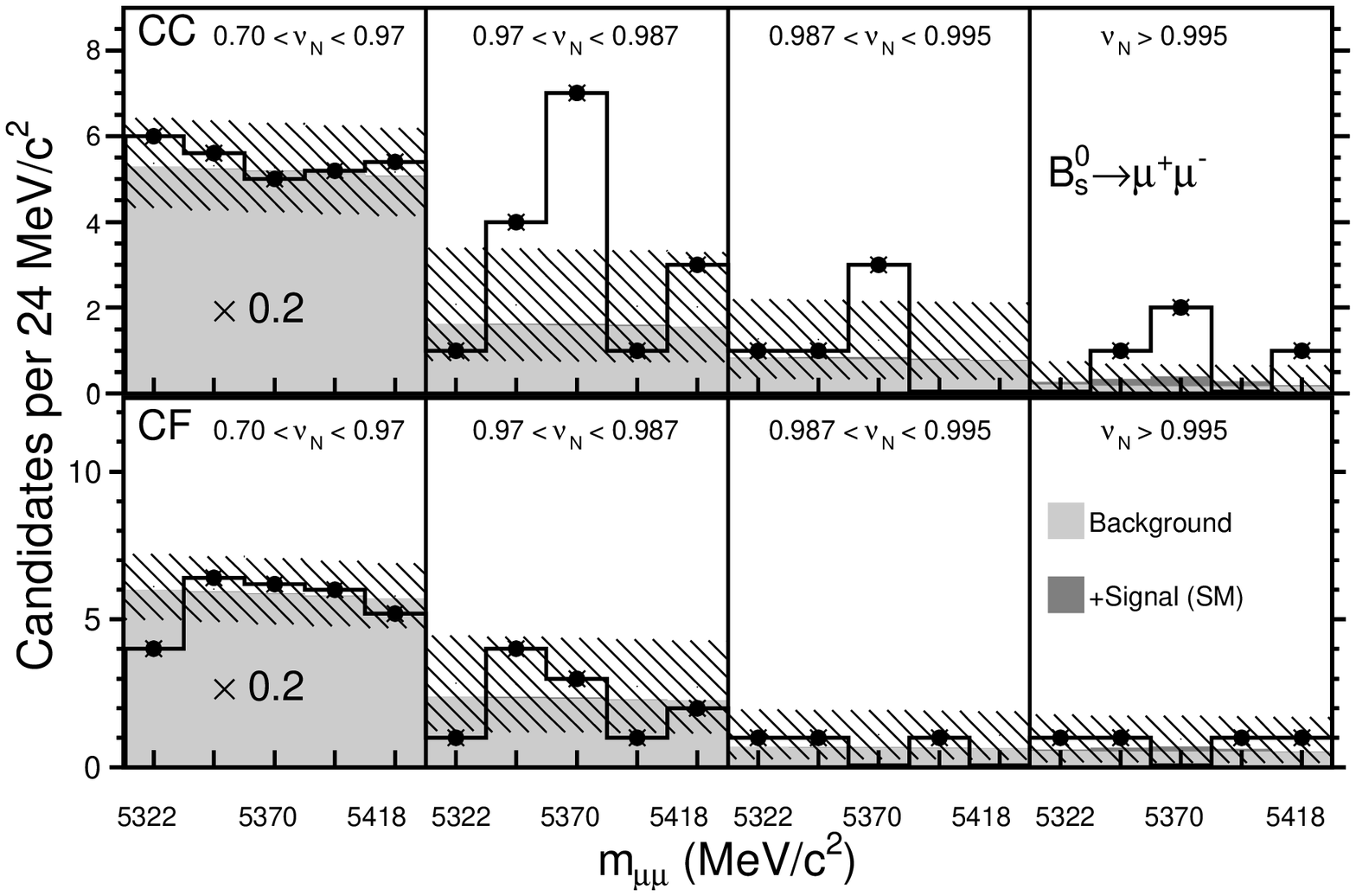}
\includegraphics[width=80mm]{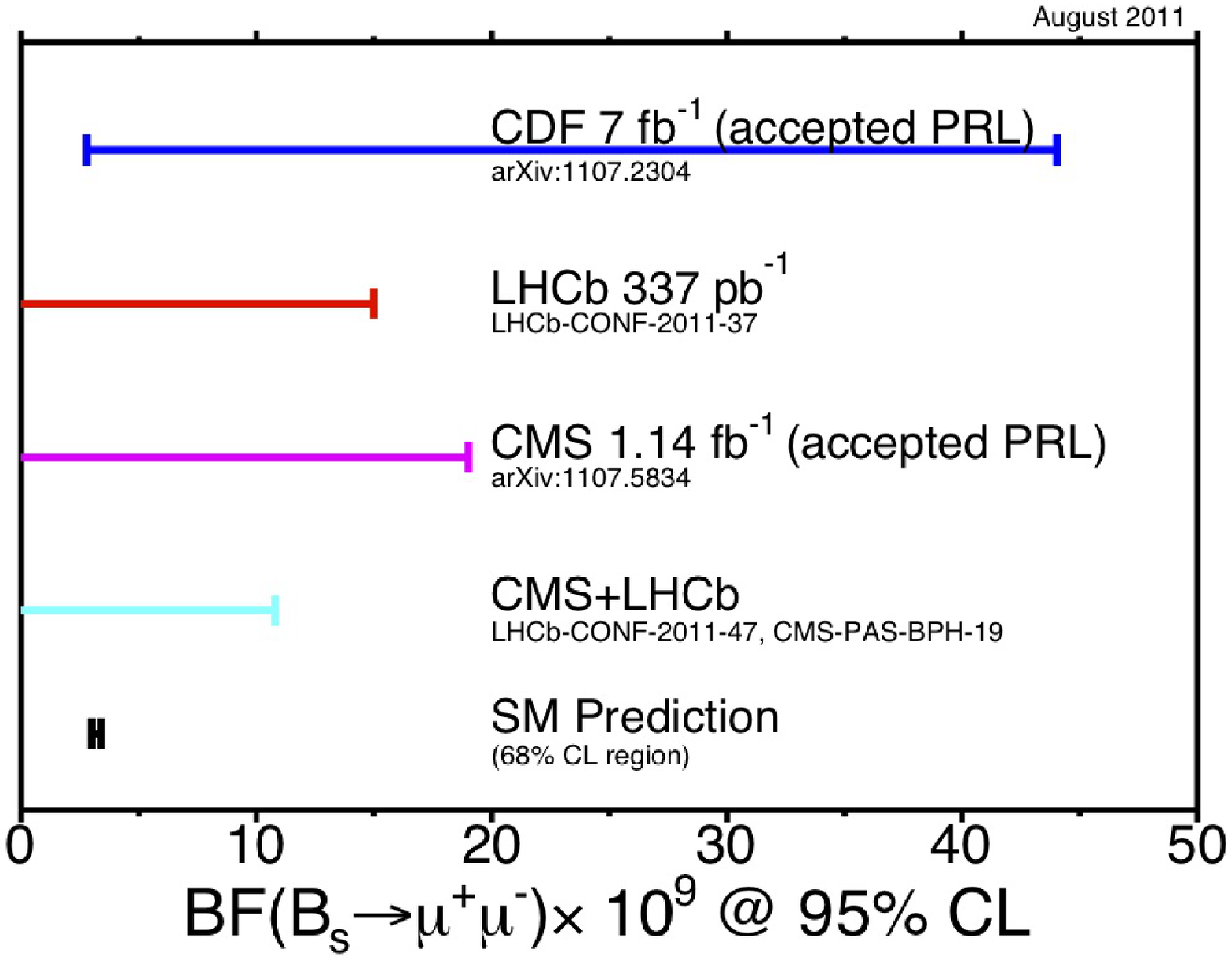}
\caption{ Comparison of background, SM prediction, and observations, separating CC and CF, but combining 5 lowest NN bins 
	in $B_s \rightarrow \mu^+ \mu^-$ decays (left). Comparison between the $B_s \rightarrow \mu^+\mu^-$ allowed branching 
	fractions at CDF, LHCb CMS and the SM prediction (right).}
\label{fig:Bs_mumu}
\end{figure}

The results for the $B_s$ region are shown in Fig.~\ref{fig:Bs_mumu}. 
There is an excess of events over background concentrated in the region 
with $\nu_{NN}$~$>~0.97$. The p-value for background-only hypothesis is 0.27\%. 
If we consider only the two highest NN bins the p-value becomes 0.66\%. 
If $B_s \rightarrow \mu^+\mu^-$ events are included in the pseudo-experiments at 
the SM level ($BR = 3.2 \times 10^{-9}$) a p-value of 1.9\% (4.1\%) is obtained using all 
(only the highest 2) NN bins. 

A log-likelihood fit is used to determine the $BR(B_s \rightarrow \mu^+\mu^-)$ most consistent with the data in 
the $B_s$ search region:
\begin{center}  
$BR(B_s \rightarrow \mu^+\mu^-) = (1.8 ^{+1.1}_{-0.9}) \times 10^{-8}$. 
\end{center} 
Additionally, 90\% C.L. bounds are set on the braching fraction: 
\begin{center}  
$4.6 \times 10^{-9} < BR(B_s \rightarrow \mu^+\mu^-) < 3.9 \times 10^{-8}$.
\end{center} 

The CDF experiment investigates the excess in the $0.97~<$~$\nu_{NN}$~$<~0.987$~bin which appears 
to be a statistical fluctuation of the background as there 
is no significant expectation of $B_s \rightarrow \mu^+\mu^-$ signal consistent with the observation 
in the two highest NN bins. 
%The source of the data excess in the $0.970 < \nu_{NN} <0.987$ range of the 
%$B_s$ signal region is investigated. 
The same events, the same fits and the same methodologies 
are used for both $B_s$ and $B^0$ searches. Since the data in the $B^0$ search region shows no excess, 
problems with the background estimates are ruled out. The only peaking background in this mass region 
is from $B \rightarrow h^+ h'^-$ decays, whose contribution to the $B^0$ search window is ten times larger 
than to the $B_s$ search window. However, no data excess is seen in the $B^0$ search window.  
The NN studies find no evidence of over-training or  
$\nu_{NN} - m_{\mu\mu}$ correlations and no evidence for mis-modeling of the $\nu_{NN}$ shape. 
The most plausible explanation for the data excess in the $0.97~<$~$\nu_{NN}$~$<~0.987$ is a 
statistical fluctuation.    

The Tevatron results are consistent with the recent measurements from the LHCb experiment,  
$BR(B_s \rightarrow \mu^+\mu^-) < 12 (15) \times 10^{-9}$ at 90 (95)\% CL~\cite{LHCb_Bs_mumu}
and the CMS experiment $BR(B_s \rightarrow \mu^+\mu^-) < 16 (19) \times 10^{-9}$ at 90 (95)\% CL~\cite{CMS_Bs_mumu}.

\section{Flavor Changing Neutral Currents in $b \rightarrow s \mu \mu$ Decays}

Rare decays of bottom hadrons mediated by the flavor-changing neutral current (FCNC) 
process $b \rightarrow s \mu \mu$ occur in the SM through higher order amplitudes. 
A variety of beyond-the-standard-model (BSM) theories, on the other hand, favor enhanced rates for 
these FCNC decays. One can obtain rich information about the $b \rightarrow s \mu \mu$ dynamics by 
measuring of the branching ratios, their dependence on the di-lepton mass distributions, 
and the angular distributions of the decay products~\cite{b_smumu_br}.
The CDF experiment has analyzed the following decays governed by the $b \rightarrow s \mu \mu$ transition:
\begin{center}
$\Lambda^0_b \rightarrow \Lambda \mu^+\mu^-$ \\
$B^0_s \rightarrow \phi \mu^+\mu^-$ \\
$B^+ \rightarrow K^+ \mu^+\mu^-$ \\
$B^0 \rightarrow K^{*0}(892) \mu^+\mu^-$ \\
$B^0 \rightarrow K^0 \mu^+\mu^-$ and \\
$B^+ \rightarrow K^{*+}(892) \mu^+\mu^-$.
\end{center}
In addition to branching fractions and differential branching fractions of these decays, the angular distributions in 
$B \rightarrow K^{(*)} \mu^+\mu^-$ decays are measured, as well. The analysis is based on a dataset corresponding to 
6.8~fb$^{-1}$ of integrated luminosity.
Previous iterations used 4.4~fb$^{-1}$ and 924~pb$^{-1}$, respectively~\cite{b_smumu_old}.

\begin{figure}[tb]
\includegraphics[width=50mm]{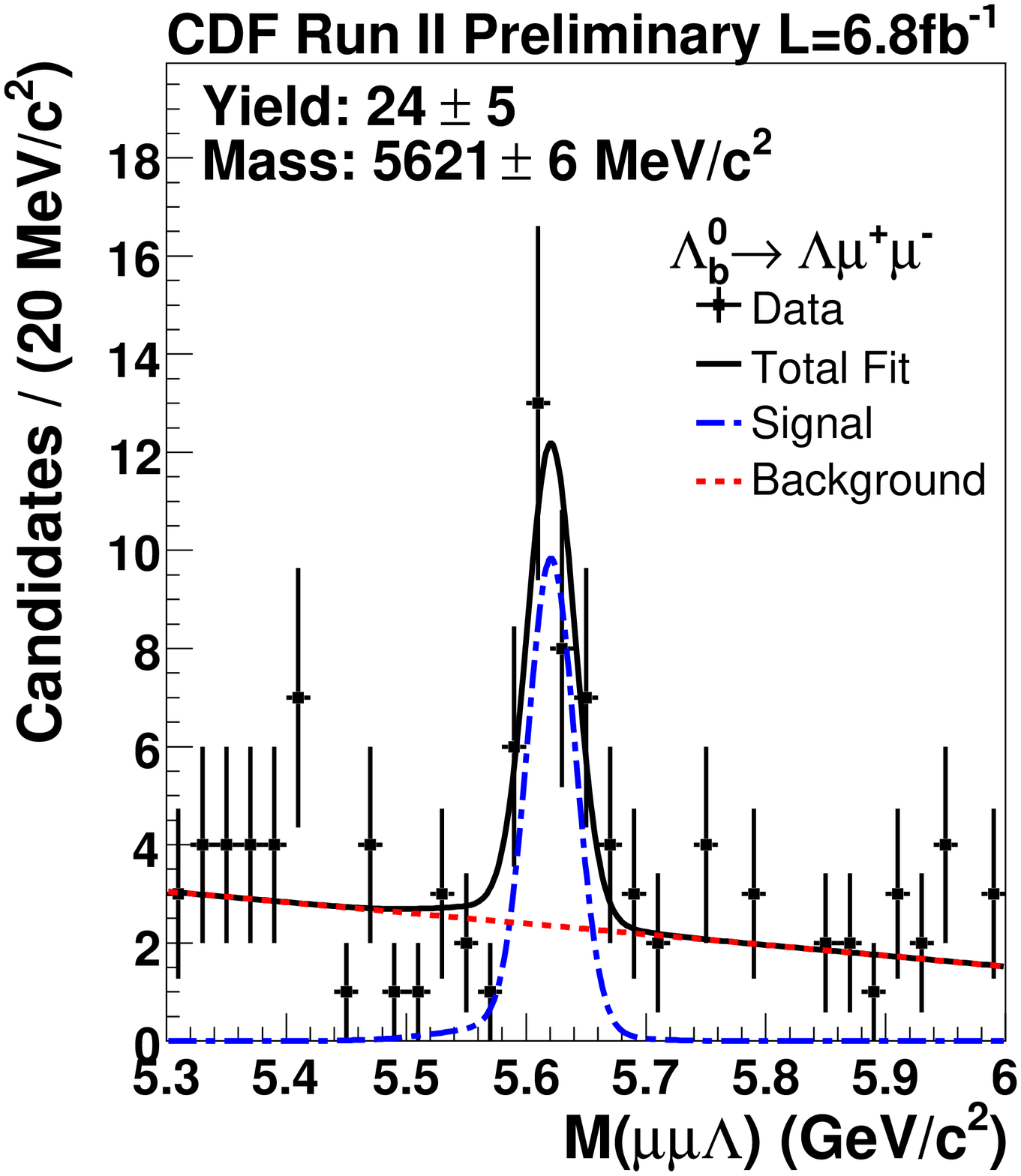}
\includegraphics[width=55mm]{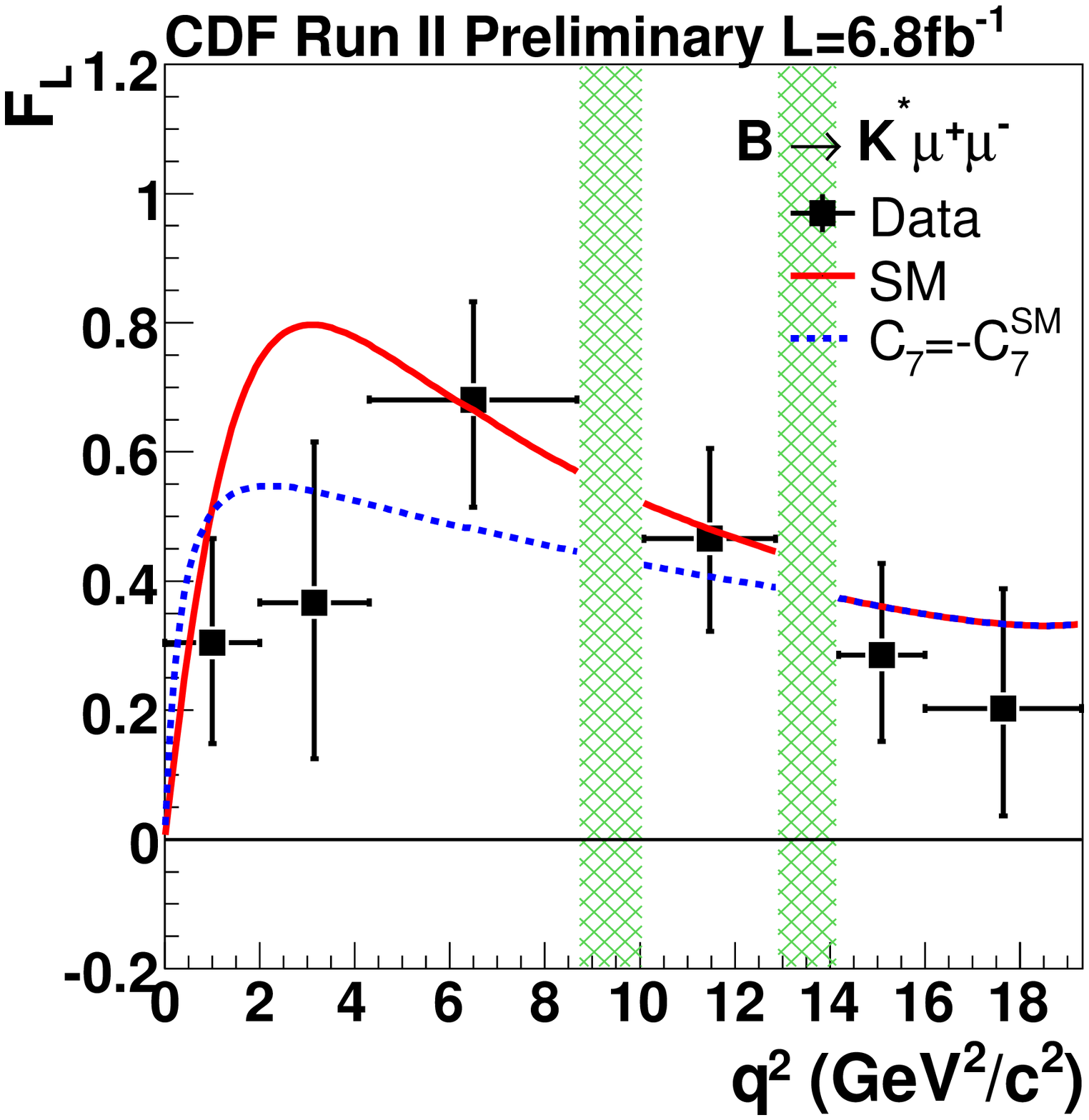}
\includegraphics[width=55mm]{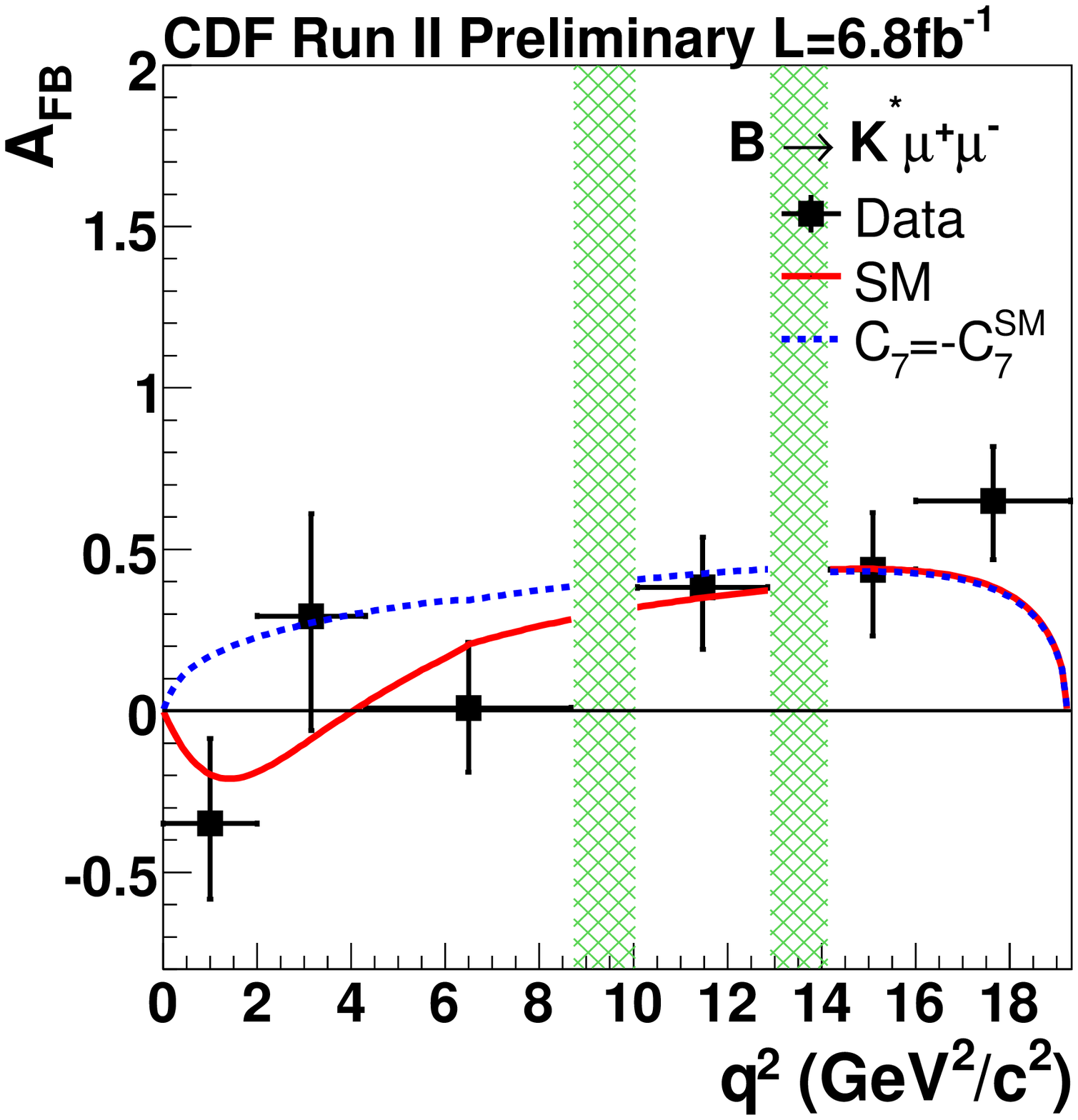}
\caption{ $\Lambda^0_b \rightarrow \Lambda \mu^+\mu^-$ candidate mass distribution; 
	the signal significance is 5.8 Gaussian sigma. (left).  
	Parameters $F_L$ (center) and $A_{FB}$ (right) as function of the di-muon invariant mass in 
	 $B \rightarrow K^* \mu^+ \mu^-$ decays (simultaneous fit of $K^{*0}$ and $K^{*+}$ channels).
	The red curves represent the SM expectations~\cite{b_smumu_sm,Bobeth_2010wg} while the blue curves correspond to 
	a supergravity model with large tan$(\beta)$~\cite{b_smumu_bsm}. }
\label{fig:Lb_Lmumu}
\end{figure}

The results include the first observation of the baryonic FCNC decay 
$\Lambda^0_b \rightarrow \Lambda \mu^+\mu^-$ and the first measurement of its branching fraction 
and of the differential branching fraction as a function of squared dimuon mass.
Fig.~\ref{fig:Lb_Lmumu} shows the invariant mass distribution of $\Lambda \mu^+\mu^-$ from $\Lambda^0_b$ decays.

Most precise branching fraction measurements in $b \rightarrow s \mu \mu$ decays are determined as follows:
\begin{center}
$BR(\Lambda^0_b \rightarrow \Lambda \mu^+\mu^-) = [1.73 \pm 0.42 (\rm stat.) \pm 0.55 (\rm syst.) ] \times 10^{-6}$ \\
$BR(B^0_s \rightarrow \phi \mu^+\mu^-) = [1.47 \pm 0.24 (\rm stat.) \pm 0.46 (\rm syst.) ] \times 10^{-6}$ \\
$BR(B^+ \rightarrow K^+ \mu^+\mu^-) = [0.46 \pm 0.04 (\rm stat.) \pm 0.02 (\rm syst.) ] \times 10^{-6}$ \\
$BR(B^0 \rightarrow K^{*0}(892) \mu^+\mu^-) = [1.02 \pm 0.10 (\rm stat.) \pm 0.06 (\rm syst.) ] \times 10^{-6}$ \\
$BR(B^0 \rightarrow K^0 \mu^+\mu^-) = [0.32 \pm 0.10 (\rm stat.) \pm 0.02 (\rm syst.) ] \times 10^{-6}$ \\
$BR(B^+ \rightarrow K^{*+}(892) \mu^+\mu^-) = [0.95 \pm 0.32 (\rm stat.) \pm 0.08 (\rm syst.) ] \times 10^{-6}$. 
\end{center}

The full differential decay distribution for the decay $B \rightarrow K^{(*)} \mu^+\mu^-$ is described by 
four independent kinematic variables: 
the di-muon invariant mass squared $q^2$, 
the angle $\theta_{\mu}$ between the muon $\mu^{+/-}$ direction and the direction opposite to the $B/{\bar B}$-meson in the di-muon rest 
frame, 
the angle $\theta_{K}$ between the kaon direction and the direction opposite to the $B$-meson in the 
$K^*$ rest frame, and the angle $\phi$ between the two planes formed by the di-muon and the $K-\pi$ systems. 
The distributions of $\theta_{\mu}$, $\theta_K$, and $\phi$ are projected from the full differential decay 
distribution and can be parametrized with four angular observables, $A_{FB}$, $F_L$, $A_T^{(2)}$ and $A_{im}$:
\begin{center}
$\frac{1}{\Gamma} \frac{d\Gamma}{d\cos(\theta_K)} = \frac{3}{2} F_L \cos^2\theta_K + \frac{3}{4} (1-F_L) (1-\cos^2\theta_K)$, \\
$\frac{1}{\Gamma} \frac{d\Gamma}{d\cos(\theta_{\mu})} = \frac{3}{4} F_L (1-\cos^2\theta_{\mu}) + 
			\frac{3}{8} (1-F_L) (1+\cos^2\theta_{\mu}) + A_{FB} \cos\theta_{\mu}$, \\
$\frac{1}{\Gamma} \frac{d\Gamma}{d\phi} = \frac{\pi}{2} [ 1+ \frac{1}{2} (1-F_L) A_T^{(2)} \cos2\phi + A_{im} \sin2\phi ] $, \\
\end{center}
where $\Gamma = \Gamma (B \rightarrow K^* \mu^+ \mu^-)$, $A_{FB}$ is the muon forward-backward asymmetry, 
$F_L$ is the $K^*$ longitudinal polarization fraction, $A_T^{(2)}$ is the transverse polarization asymmetry, 
and $A_{im}$ is the T-odd CP asymmetry of the transverse polarizations. 
Fig.~\ref{fig:Lb_Lmumu} shows agreement between the $F_L$, $A_{FB}$ and $A_T^{(2)}$ as function of the di-muon 
invariant mass and the SM expectations~\cite{b_smumu_sm,Bobeth_2010wg}.
The angular analysis results are among the most precise measurements to date. The right-handed current  
sensitive observables $A_T^{(2)}$ and $A_{im}$ are measured for the first time.

\section{Like-Sign Dimuon Asymmetry}    

The D0 collaboration presents an updated measurement~\cite{dimuon_asym} of the like-sign dimuon charge 
asymmetry in semi-leptonic decays of $b$-hadrons using a data sample corresponding to 
9~fb$^{-1}$ of integrated luminosity. The like-sign dimuon asymmetry is defined as: 
\begin{center}
$A_{sl}^b = \frac{N_b^{++} - N_b^{--}}{N_b^{++} + N_b^{--}} = C_d a_{sl}^d + C_s a_{sl}^s$, 
\end{center} 
where $N_b^{++}$ and $N_b^{--}$ are the number of events containing two muons of same charge, 
produced in semi-leptonic decays of $b$-hadrons. The asymmetries $a_{sl}^q$, where $q = s/d$,  
are defined as:
\begin{center}
$a_{sl}^q = \frac{\Gamma(\bar{B}_q^0 \rightarrow \ \mu^+ X) - \Gamma(B_q^0 \rightarrow \ \mu^- X)}{\Gamma(\bar{B}_q^0 \rightarrow \ \mu^+ X) + \Gamma(B_q^0 \rightarrow \ \mu^- X)} = \frac{\Delta \Gamma_q}{\Delta M_q}$ tan$(\phi_q)$, 
\end{center}
where $\phi_q$ is a CP violating phase and $\Delta M_q$ and $\Delta \Gamma_q$ are the mass and width difference between 
the eigenstates of the time propagation operator of the neutral $B^0_q$ systems. The coefficients $C_d$ and $C_s$ 
depend on the mean mixing probabilities and on the production rates of the $B^0$ and $B_s^0$ mesons~\cite{HFAG}: 
\begin{center}
$C_d = 0.594 \pm 0.022$ and $C_s = 0.406 \pm 0.022$.
\end{center}
Using the SM predictions for $a_{sl}^q$~\cite{aslq_th}, one finds 
\begin{center}
$A_{sl}^b(SM) = (-0.028^{+0.005}_{-0.006})\%$. 
\end{center}
This theoretical uncertainties are negligible compared to the experimental sensitivity. 
A previous D0 analysis based on 6.1~fb$^{-1}$ of integrated luminosity~\cite{dimuon_asym_6fb} 
revealed a dimuon asymmetry of 3.2 standard deviations away from the SM expectation: 
\begin{center}
$A_{sl}^b = (-0.00957 \pm 0.00251 (\rm stat.) \pm 0.00146 (\rm syst.)$. 
\end{center}
The updated measurement not only uses an increased dataset due to increased integrated 
luminosity from 6.1~fb$^{-1}$ to 9~fb$^{-1}$, but also includes analysis improvements: 
13\% increase in data sample due to looser muon longitudinal momentum selection and 
20\% reduction in kaon and pion decay-in-flight backgrounds.
In addition, muon impact parameter studies support the hypothesis that muons are indeed 
from $B$ decays. The new result is 3.9 standard deviations from the SM expectation:  
\begin{center}
$A_{sl}^b = (-0.787 \pm 0.172 (\rm stat.) \pm 0.093 (\rm syst.) )\%$.
\end{center} 
and it represents one of the most interesting high energy physics results.

\begin{figure}[tb]
\includegraphics[width=65mm]{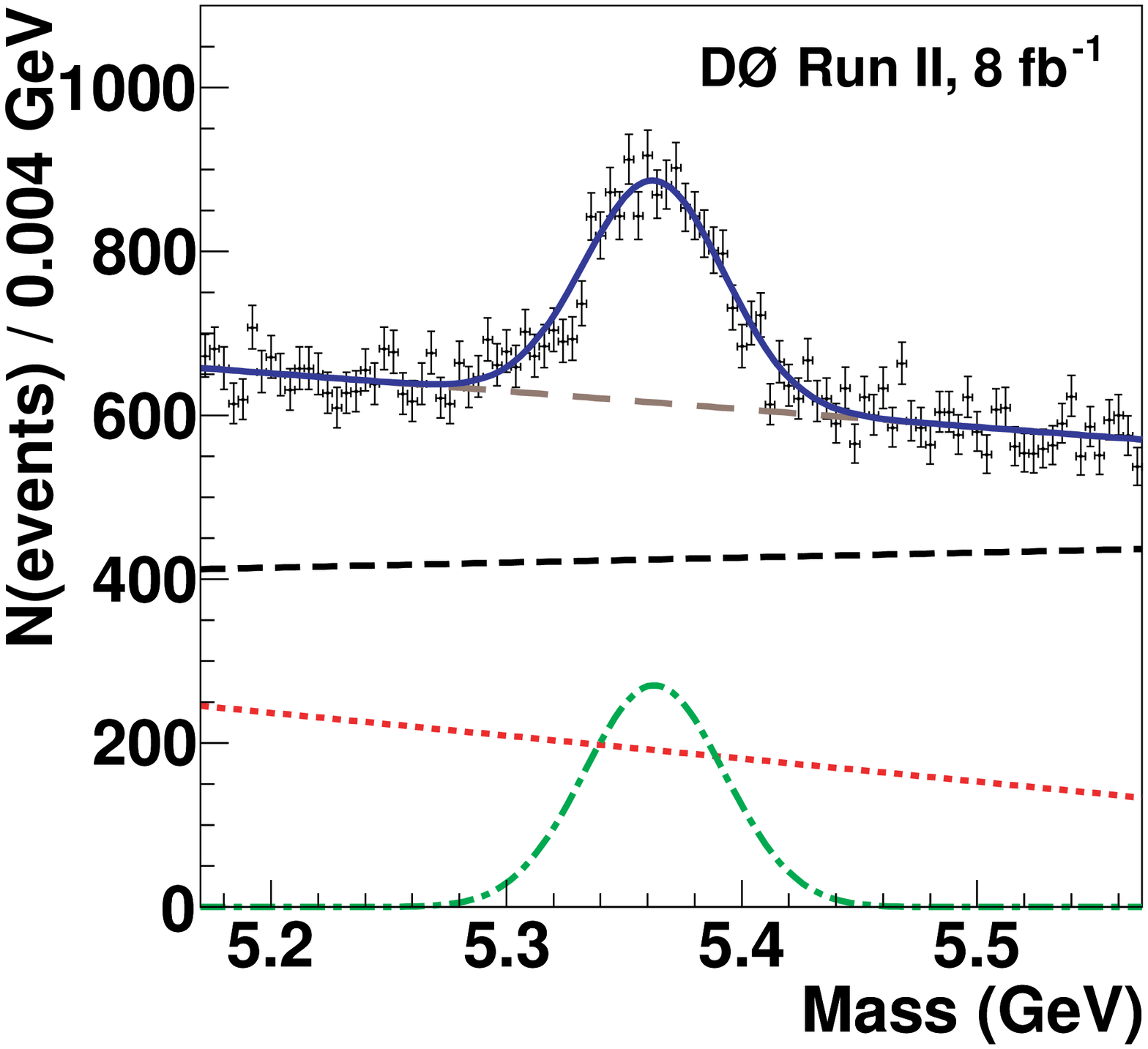}
\hskip1cm
\includegraphics[width=90mm]{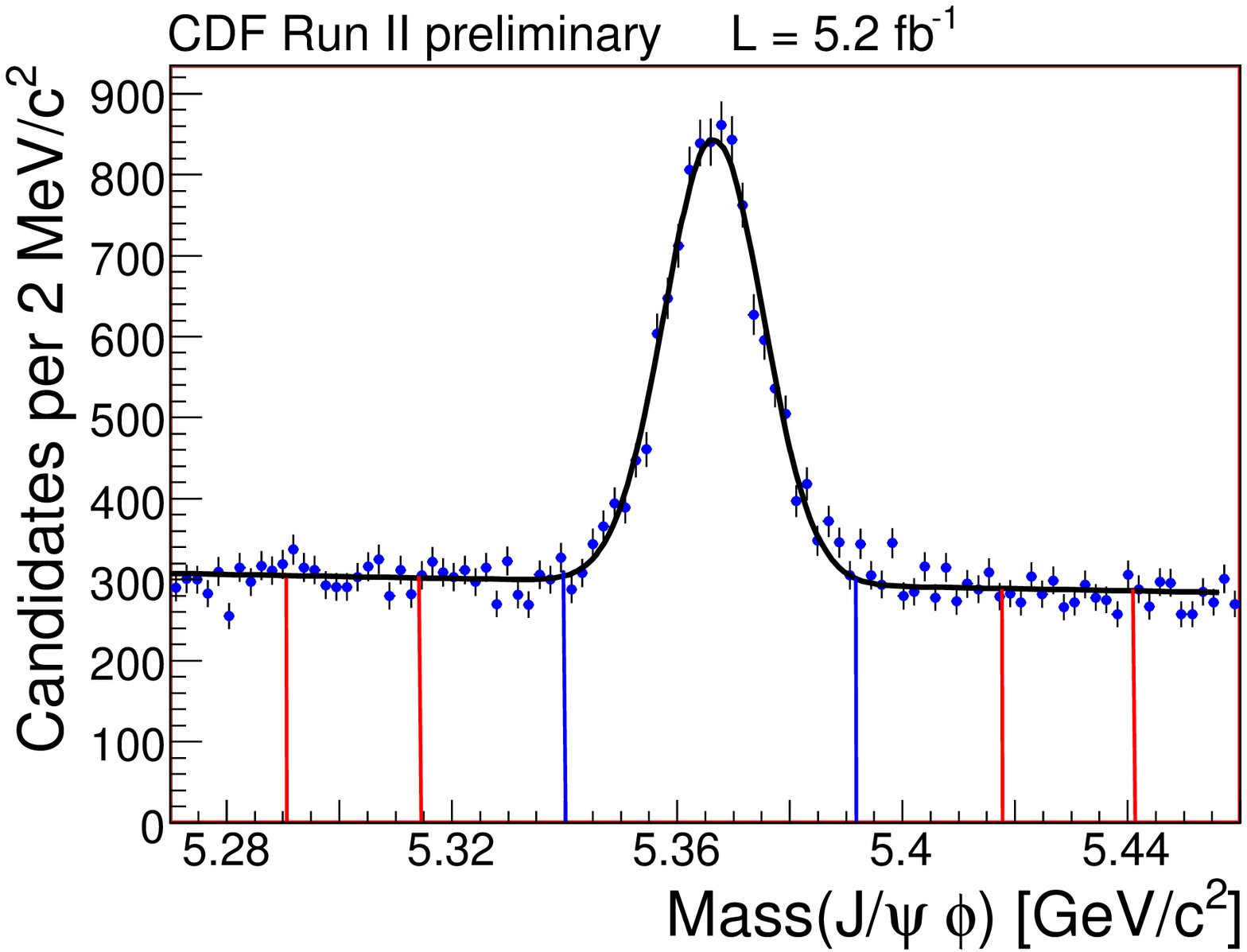}
\caption{ $K^+K^-\mu^+\mu^-$ invariant mass distribution from $B_s \rightarrow J/\psi \phi$ candidates at D0 (left) and CDF (right).  }
\label{bs_mass}
\end{figure}

\section {CP Violation in $B_s \rightarrow J/\psi \phi$ Decays }

While CP violation has been well-measured and found to agree with the SM expectations 
in kaon and in most $B$-meson decays, the study 
of CP violation in decays of $B_s$ mesons is still in its early stages, with the 
first results from $B_s \rightarrow J/\psi \phi$ decays reported by the CDF and D0 
collaborations in the last couple of years~\cite{cdf_beta_s_prl,paulini,d0_beta_s_prl}. 
In these decays, CP violation occurs through the interference between the decay amplitudes 
with and without mixing. In the SM the relative phase between the decay amplitudes with and 
without mixing is $\beta_s^{SM}={\rm arg}(-V_{ts}V_{tb}^{*}/V_{cs}V_{cb}^{*})$ and it 
is expected to be very small~\cite{bigi-sanda,Ref:lenz}. New physics contributions 
manifested in the $B_s^0$ mixing amplitude may alter this mixing phase by a 
quantity $\phi_s^{NP}$ leading to an observed mixing phase 
$2\beta_s^{J/\psi\,\phi} = 2\beta_s^{SM} - \phi_s^{NP}$. 
Large values of the observed $\beta_s^{J/\psi\,\phi}$ would be an indication of physics 
beyond the SM~\cite{Ref:lenz,Ref:hou,Ref:ligeti,theory}.        
It is interesting to note that certain SUSY models with large tan($\beta$) predict 
enhanced $BR(B_s \rightarrow \mu \mu)$ for large CP violating mixing phase in 
$B_s \rightarrow J/\psi \phi$ decays~\cite{Altmannshofer}.   

Early measurements of the CP violation parameter $\beta_s$ from the CDF~\cite{paulini} 
and D0~\cite{d0_beta_s_prl} collaborations showed 
small deviations from the SM~\cite{bigi-sanda,Ref:lenz}, however, a combination~\cite{cdf_d0_comb}  
of CDF and D0 analyses, based on 2.8~fb$^{-1}$ of integrated luminosity, revealed 
a deviation of slightly more that two standard deviations with respect to the SM 
predictions. 
More recent updates of this measurements were performed by both the CDF~\cite{cdf_bs_5.2fb} 
and the D0~\cite{d0_bs_8.0fb} experiments using data samples corresponding to 5.2~fb$^{-1}$ 
and 8.0~fb$^{-1}$ of integrated luminosity, respectively. The CDF experiment 
has a $B_s$ yield of $\approx 6500$ signal events, while the D0 experiment reports a yield 
of $\approx 5500$ signal events, as shown in Fig.~\ref{bs_mass}. 

\begin{figure}[tb]
\includegraphics[width=90mm]{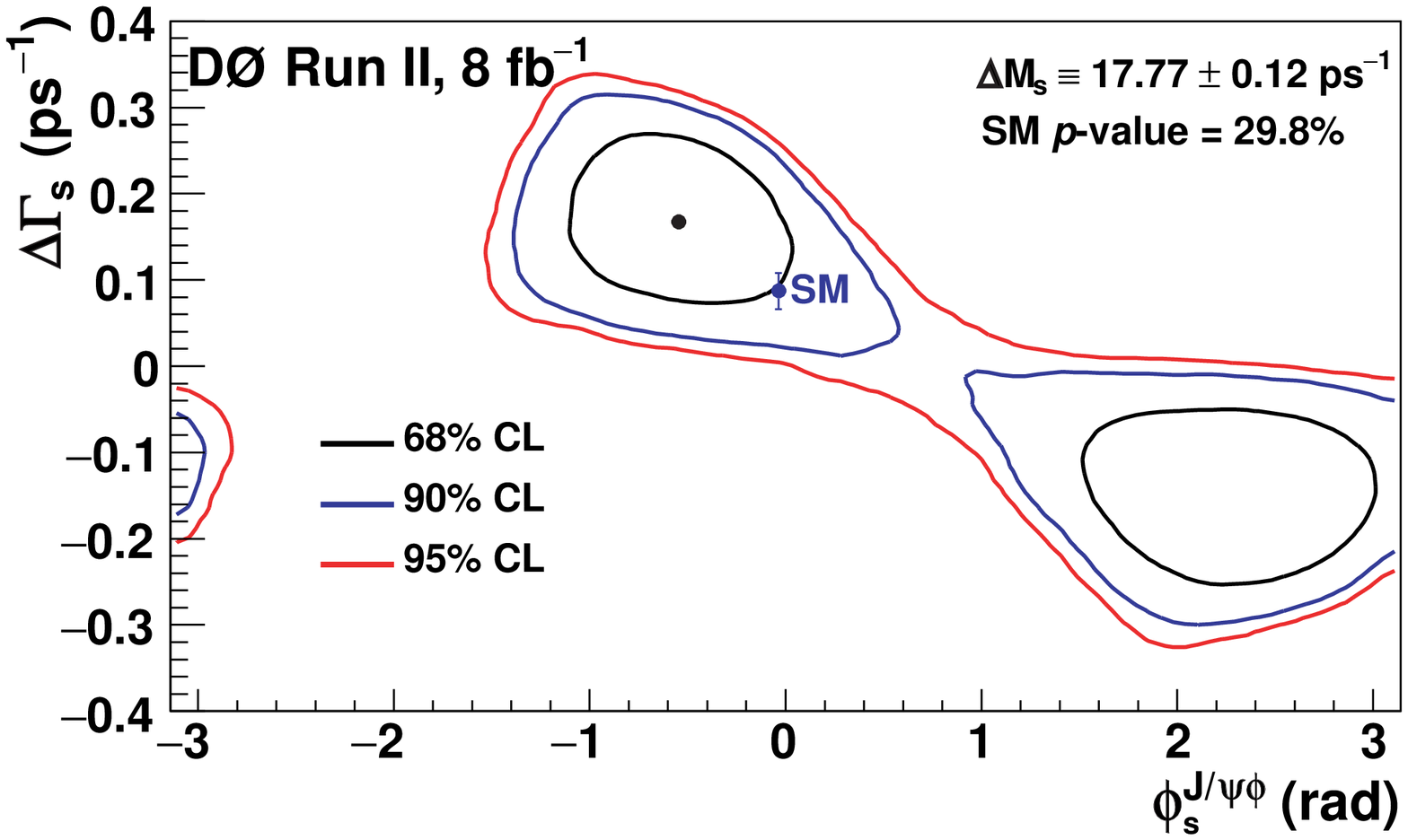}
\hskip1cm
\includegraphics[width=65mm]{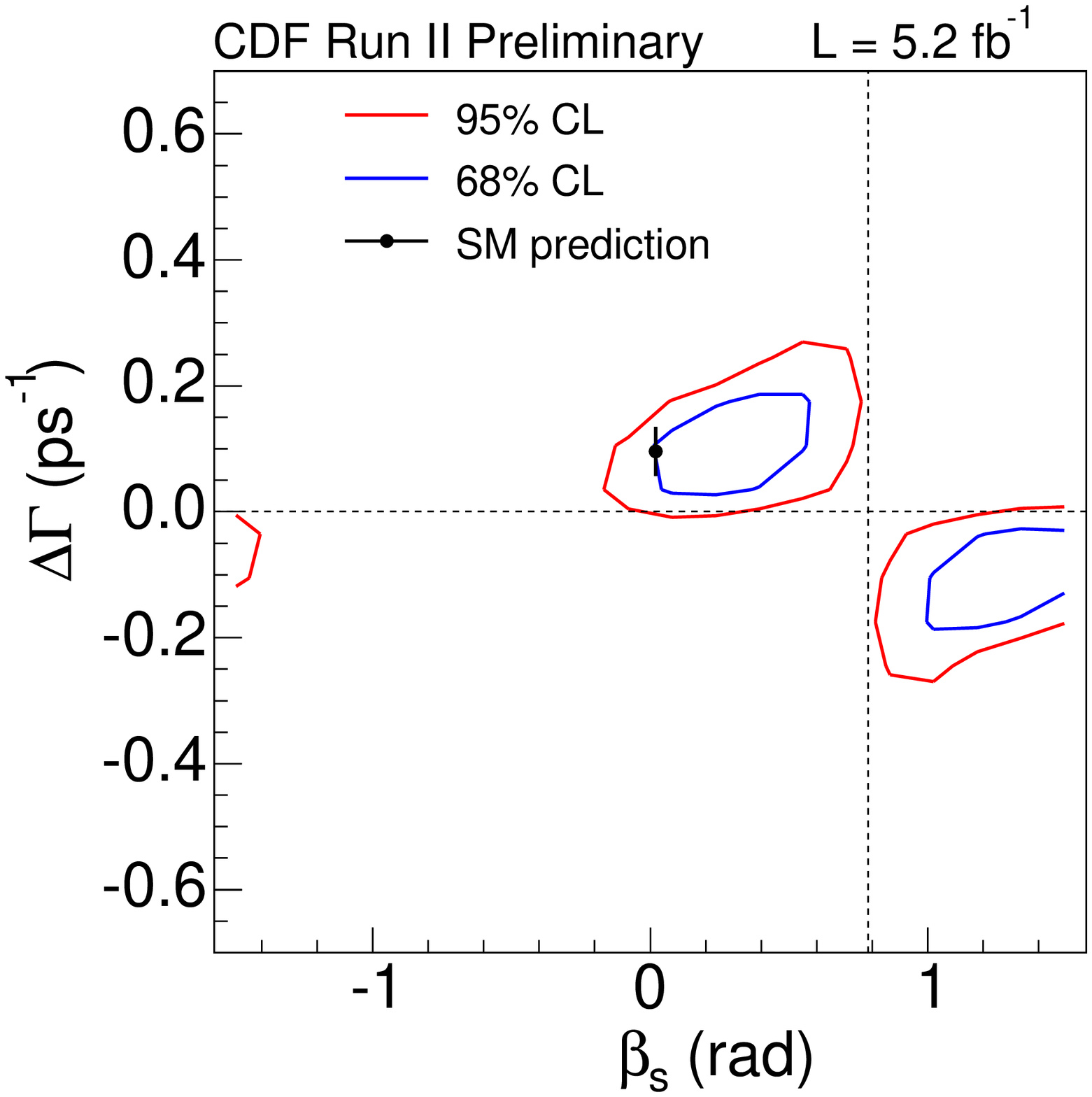}
\caption{ Confidence regions in the $\phi_s-\Delta \Gamma_s$ plane from D0 (left) and 
	in the $\beta_s-\Delta \Gamma_s$ plane from CDF (right). When large CP violation effects 
	from new physics are present, $\phi_s$ and $\beta_s$ are related by the simple 
	equation $\phi_s = -2 \beta_s$. }
\label{bs_dg_contours}
\end{figure}

The updated measurements show better agreement with the SM expectation. The deviations are 
at one standard deviation level for each experiment. 
The CDF experiment finds~\cite{cdf_bs_5.2fb} that the CP violation phase $\beta_s$ is within the ranges 
[0.02,0.52]~U~[1.08,1.55]~radians at the 68\% CL. The corresponding D0 result is~\cite{d0_bs_8.0fb} 
$\beta_s = 0.28^{+0.19}_{-0.18}$~radians or $\phi_s = -2\beta_s = -0.55^{+0.38}_{-0.36}$~radians. 
The two dimensional confidence regions in the $\beta_s-\Delta\Gamma_s$ plane are shown 
in Fig.~\ref{bs_dg_contours}. 

Apart from increasing the sample sizes, each experiment includes in the analysis the s-wave contribution 
from $B_s \rightarrow J/\psi K^+K^-$, where the $K^+K^-$ pair is in an s-wave state. The s-wave could be 
either the $f_0(980)$ state or a non-resonant $K^+K^-$ state. The s-wave contribution to the CDF analysis 
is found to be less than 6.7\% at the 95\% CL while the corresponding D0 fraction is $(17.3 \pm 3.6)\%$. 
The difference between the two s-wave contributions in the two analyses is still to be understood. 

The $B_s$ mean lifetime, $\tau_s$, the decay with difference between the $B_s$ mass eigenstates, 
$\Delta \Gamma_s$, the polarization fractions in the transversity basis $|A_{0}(0)|^2$ 
and $|A_{\parallel}(0)|^2$ and the strong phases 
$\varphi_{\parallel} = {\rm arg}(A_{\parallel}(0) A_0^*(0))$, 
 $\varphi_{\perp}     = {\rm arg}(A_{\perp}(0) A_0^*(0))$ and 
 $\varphi_s          = {\rm arg}(A_s(0) A_0^*(0))$ are investigated. The CDF results are: 
\begin{center}
$c\tau_s = 458.6 \pm 7.6 (\rm stat.) \pm 3.6 (\rm syst.) \mu$m \\
$\Delta \Gamma_s = 0.075 \pm 0.035 (\rm stat.) \pm 0.01 (\rm syst.)$~ps$^{-1}$ \\ 
$|A_{||}(0)|^2 = 0.231 \pm 0.014 (\rm stat.) \pm 0.015 (\rm syst.)$ \\
$|A_{0}(0)|^2  = 0.524 \pm 0.013 (\rm stat.) \pm 0.015 (\rm syst.)$ \\
$\varphi_{\perp}  = 2.95  \pm 0.64  (\rm stat.) \pm 0.07  (\rm syst.)$ 
\end{center}
Due to the low s-wave fraction present in the CDF data, the analysis has no sensitivity 
to the relative phase $\varphi_s$ between the s-wave amplitude $A_s(0)$ 
and the amplitude $A_0(0)$. The uncertainties 
of the phase $\varphi_{\parallel}$ are still being investigated.   
The corresponding D0 results, including both statistic and systematic uncertainties, are:
\begin{center}
$c\tau_s = 1.443^{+0.038}_{-0.035}$~ps \\
$\Delta \Gamma_s = 0.163^{+0.065}_{-0.064} $~ps$^{-1}$ \\ 
$|A_{||}(0)|^2 = 0.231^{+0.024}_{-0.030}$ \\
$|A_{0}(0)|^2  = 0.558^{+0.017}_{-0.019}$ \\
$\delta_{\parallel} = 3.15 \pm 0.22$ \\
$\cos(\delta_{\perp}-\delta_s) = 0.11^{+0.27}_{-0.25}$ 
\end{center}

The Tevatron experiments have pioneered the exploration
of CP violation in the neutral $B_s$ system. They have strongly constrained the size of 
possible presence of new physics contributions and will further restrict it with
the full Run II data sample. Since the time of this presentation, the LHCb experiment 
has presented~\cite{beta_s_lhcb} an updated analysis of $B_s \rightarrow J/\psi \phi$ decays, 
providing competitive uncertainties on the CP violation parameter $\beta_s$, which was found 
to be in agreement with the Tevatron results and also with the SM prediction. 

\section{Study of $B_s \rightarrow J/\psi f_0(980)$ Decays}

Due to the small SM value of the phase 
$\phi_s = {\rm arg}(-M^s_{12}/\Gamma^s_{12}) = (4.2 \pm 1.4) \times 10^{-3}$~radians~\cite{Ref:lenz}, 
the $B_s$ mass eigenstates and the CP eigenstates coincide to a good approximation.   
Here $M^s_{12}$ and $\Gamma^s_{12}$ are the off-diagonal elements of the mass and decay 
matrices which describe the time evolution of the neutral $B_s$ system. 
The measurement of the mean $B_s$ lifetime decaying to a CP eigenstate 
provides directly the lifetime of the corresponding mass eigenstate. 
If new physics has large contributions to $\phi_s$, then the the mass 
and CP eigenstates are no longer the same. In this case, the measured lifetime 
corresponds to the weighted average of the lifetimes of the two mass eigenstates, 
with weights depending on the size of the CP violating phase $\phi_s$~\cite{theory}. 
The measurement of the $B_s$ lifetime in a final state which is a CP eigenstate 
provides constraints on the width difference, $\Delta \Gamma_s$, and on 
the CP violating phase in $B_s$ mixing, $\phi_s$~\cite{Fleischer_Knegjens,Fazio}.

Since the final state in the decay $B_s \rightarrow J/\psi f_0(980)$ with 
$f_0 \rightarrow \pi^+ \pi^-$ is a CP eigenstate,  
this decay can be used to measure the CP violating phase $\beta_s = {\rm arg}[(-V_{ts}V^*_{tb})/(V_{cs}V^*_{cb})]$ 
without performing an angular analysis~\cite{stone}. In case of large CP violation new physics effects in mixing, 
it holds that $\phi_s \simeq -2\beta_s$. 
A measurement of the phase $\beta_s$ in $B_s \rightarrow J/\psi f_0(980), f_0 \rightarrow \pi^+ \pi^-$ 
decays was already performed by the LHCb experiment~\cite{beta_s_lhcb}.
Further interest in the decay  $B_s \rightarrow J/\psi f_0(980)$ with $f_0 \rightarrow K^+ K^-$, 
is generated by the possibility of solving the $\beta_s$ ambiguity by using the interference between 
the p-wave in $B_s \rightarrow J/\psi \phi$ decays and the s-wave in $B_s \rightarrow J/\psi f_0(980)$ 
decays.   

\begin{figure}[tb]
\includegraphics[width=75mm]{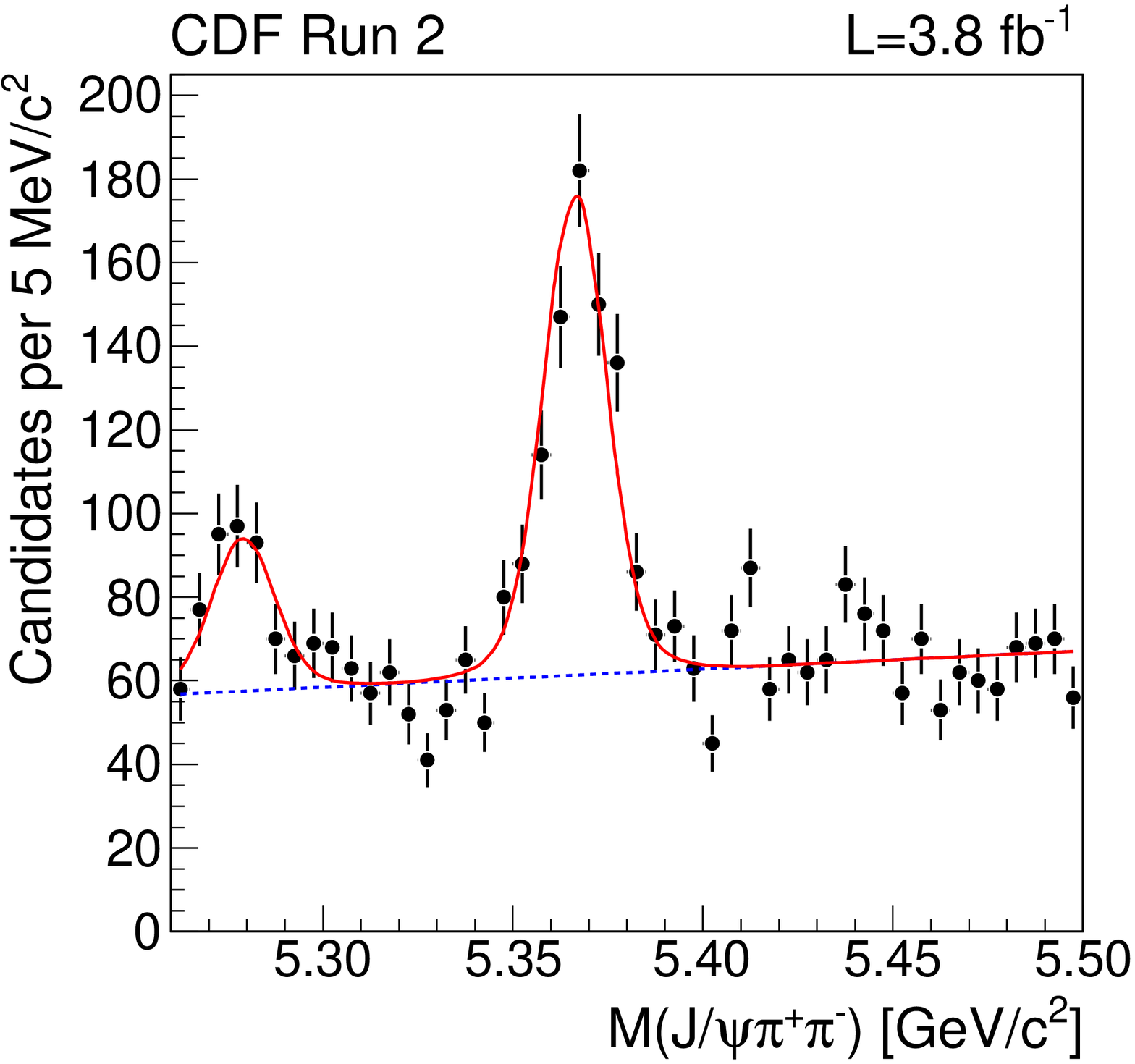}
\hskip1cm
\includegraphics[width=75mm]{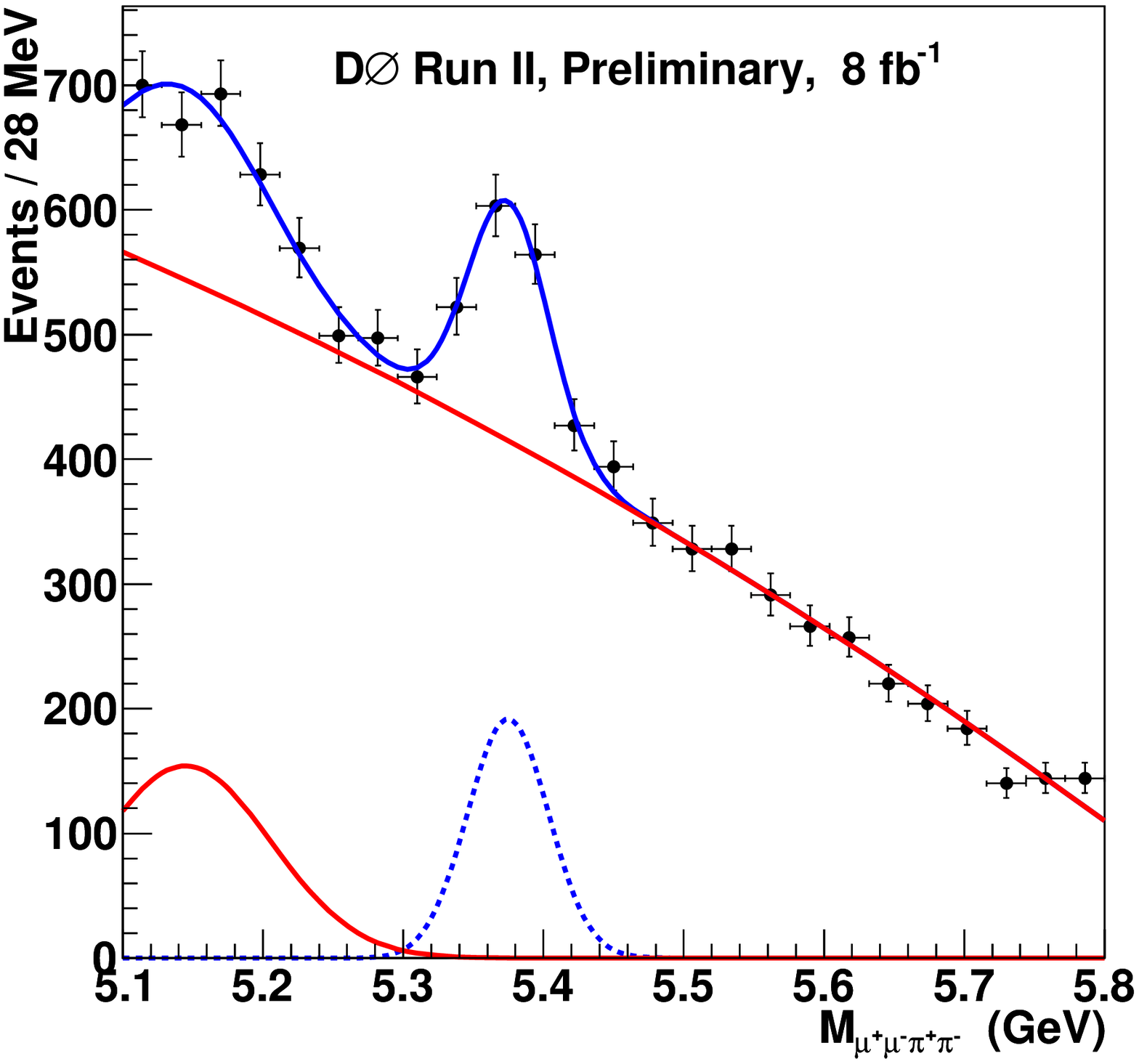}
\caption{ Invariant mass of $\mu^+\mu^-\pi^+\pi^-$ from $B_s \rightarrow J/\psi f_0(980)$, $f_0(980) \rightarrow \pi^+ \pi^-$ candidates  
	from the CDF experiment (left) and from the D0 experiment (right).}
\label{fig:cdf_f0}
\end{figure}

With a sample of 3.8~fb$^{-1}$ containing $502 \pm 37(\rm stat.) \pm 18(\rm syst.)$ signal events, 
the CDF experiment measures~\cite{cdf_f0}
\begin{center}
$R_{f_0/\phi} = \frac{BF(B_s \rightarrow J/\psi f_0(980)) \times BF(f_0(980) \rightarrow \pi^+ \pi^-)}{BF(B_s \rightarrow J/\psi \phi) \times BF(\phi \rightarrow K^+K^-)} = 0.257 \pm 0.020 (\rm stat.) \pm 0.014(\rm syst.)$,
\end{center} 
from which 
\begin{center}
$BF(B_s \rightarrow J/\psi f_0(980)) \times BF(f_0(980) \rightarrow \pi^+ \pi^-) = (1.63 \pm 0.12(\rm stat.) \pm 0.09(sys) \pm 0.50(\rm PDG)) \times 10^{-4}$ 
\end{center} 
is derived. This is the most precise determination of $R_{f_0/\phi}$ to date. 
The corresponding D0 measurements, using 8~fb$^{-1}$ of integrated luminosity, with $498 \pm 74$ signal candidates, 
yields a relative branching fraction of 
\begin{center}
$R_{f_0/\phi} = \frac{BF(B_s \rightarrow J/\psi f_0(980)) BF(f_0(980) \rightarrow \pi^+ \pi^-)}{BF(B_s \rightarrow J/\psi \phi) BF(\phi \rightarrow K^+K^-)} = 0.210 \pm 0.032 (\rm stat.) \pm 0.036(\rm syst.)$. 
\end{center}
Fig.~\ref{fig:cdf_f0} shows the CDF and D0 $\mu^+\mu^-\pi^+\pi^-$  
invariant mass from $B_s \rightarrow J/\psi f_0(980)$, $f_0(980) \rightarrow \pi^+ \pi^-$ candidates.
Both CDF and D0 results are in good agreement with the results from LHCb  
$R_{f_0/\phi} = 0.252^{+0.046}_{-0.032}(\rm stat.)^{+0.027}_{-0.033}(\rm syst.)$~\cite{lhcb_f0}  
and from Belle $R_{f_0/\phi} = 0.206^{+0.055}_{-0.034}(\rm stat.) \pm 0.052 (\rm syst.)$~\cite{belle_f0} .

In addition to the relative branching fraction, $R_{f_0/\phi}$, the CDF experiment also measured 
the mean lifetime of the $B_s$ meson in $B_s \rightarrow J/\psi f_0(980)$ decays,   
$\tau(B_s \rightarrow J/\psi f_0(980)) = 1.70^{-0.11}_{+0.12}(\rm stat.) \pm 0.03(\rm syst.)$~ps.
This result is in good agreement with theoretical expectations as well as with other determinations 
of the $B_s^H$ lifetime. 

\section{Branching Fraction, Polarization and CP Violation in $B_s \rightarrow \phi \phi$ Decays}

Studies of charmless $B_s \rightarrow \phi \phi$ decays were first performed by the CDF experiment. 
We present first measurements of the branching 
ratio, of the polarization fractions and a search 
for CP Violation~\cite{CDF_phiphi_pol} in these decays using data corresponding 
to 2.9~fb$^{-1}$ of integrated luminosity. 

Charmless $B_s$ decays are still to be fully understood. They offer the possibility 
to test our current theoretical understanding and represent promising ways to search 
for physics beyond the Standard Model. 
The $B_s \rightarrow \phi \phi$ decay is part of the so-called $B \rightarrow VV$ family 
in which the initial state $B$-meson is a pseudo-scalar 
(spin 0) and the final state $VV$ contains two vector 
mesons (spin 1). 
In particular the final state for the $B_s$ to $\phi \phi$ decay is a
superposition of CP eigenstates depending on the orbital angular momenta
of the two $\phi$ mesons.
Such decays can be used to measure the $B_s$ decay width difference 
($\Delta \Gamma_s$) and the phase responsible for CP violation in the interference 
between decays with and without mixing. 
To conserve the total angular momentum in $B_s \rightarrow \phi \phi$ decays, the 
relative orbital angular momentum between the two $\phi$ mesons in the final state must be either 0, 
1 or 2. In the angular momentum space, there are various bases which can be used to 
analyze decays of pseudo-scalars to two vector mesons, but any formalism involves 
three independent amplitudes for the three different polarizations of the decay products 
in the final state. Measuring the polarization fractions amounts to an important 
test of the corresponding theoretical predictions.  

Within the SM, the dominant process that contributes to the 
$B_s \rightarrow \phi \phi$ decay is the $b \rightarrow s \bar{s} s$ penguin 
digram. The same penguin amplitude appears in other $B \rightarrow VV$ processes 
which exhibit significant discrepancies between the measured polarization fractions and 
the SM predictions. Explanations involving both new physics scenarios as well as 
newly accounted SM effects have been suggested to explain the observations. However, none of the 
existing scenarios is convincing enough. To solve this ``polarization puzzle'' it is 
important to study as many $B \rightarrow VV$ decays as available. 
The first polarization analysis of $B_s \rightarrow \phi \phi$ decays, performed by the 
CDF experiment is presented here together with an updated measurement of the 
$B_s \rightarrow \phi \phi$ branching fraction. The $B_s \rightarrow \phi \phi$ invariant mass 
distribution is shown in Fig.~\ref{fig_bs_phi_phi_tp}.
The ratio of branching fractions is determined:

\begin{center}
$\frac{BR(B_s \rightarrow \phi\phi)}{BR(B_s \rightarrow J/\psi \phi)} = [1.78 \pm 0.14 (\rm stat.) \pm 0.20 (\rm syst.)] \times 10^{-2}$
\end{center}   

Using the experimental value of the $B_s \rightarrow J/\psi \phi$ branching ratio we obtain:

\begin{center}
   $BR(B_s \rightarrow \phi\phi) = [2.32 \pm 0.18(\rm stat.) \pm 0.26 (\rm syst.) \pm 0.78 (\rm br)] \times 10^{-5}$, 
\end{center}  
using the $BR(B_s \rightarrow J/\psi \phi)$ from~\cite{bs_jpsiphi_br}, which contributes the dominant 
uncertainty, labeled (br).
This result is compatible with the initial observation~\cite{Bs_to_phiphi_180pb}, with substantial 
improvement on the statistical uncertainty. The result is also compatible with recent theoretical 
calculations~\cite{TH_br1} and~\cite{TH_br2}.

The polarization fractions and the strong phase $\delta_{||} = A_{||}A^*_{0}$ are measured as:

\begin{center}
 $|A_0|^2 = 0.348 \pm 0.041 (\rm stat.) \pm 0.021 (\rm syst.)$ \\ 
 $|A_{||}|^2 = 0.287 \pm 0.043 (\rm stat.) \pm 0.011 (\rm syst.)$ \\  
 $|A_{\perp}|^2 = 0.365 \pm 0.044 (\rm stat.) \pm 0.027 (\rm syst.)$ \\  
 cos$(\delta_{||}) = -0.91^{+0.15}_{-0.13} (\rm stat.) \pm 0.09 (\rm syst.)$   
\end{center}

The longitudinal and transverse polarization fractions are:

\begin{center}
$ f_L = 0.348 \pm 0.041 (\rm stat.) \pm 0.021 (\rm syst.), $ \\  
$ f_T = 0.652 \pm 0.041 (\rm stat.) \pm 0.021 (\rm syst.)  $
\end{center}
 
It is clear from this measurement that the SM expected amplitude hierarchy 
$|A_0| \gg |A_{||}| \simeq |A_{\perp}|$ is not valid in $B_s \rightarrow \phi \phi$ 
decays. Instead, the observed relation between the polarization amplitudes 
is given by: $|A_0| \simeq |A_{||}| \gtrsim |A_{\perp}|$, which is similar 
to the measurements for the $\bar b \rightarrow \bar s$ penguin transition 
of $B \rightarrow \phi K^*$ decays~\cite{ct8} which were the 
origin of the polarization puzzle.

The results are compared with various theoretical predictions of the polarization 
amplitudes. We find that the central values are consistent within the uncertainty 
ranges with the expectations of QCD factorization~\cite{TH_br1}{ and with~\cite{HaiYangCheng_new}, 
while they are not in good agreement with the expectation of perturbative QCD~\cite{TH_br2}.

Although the $B_s \rightarrow \phi \phi$ data sample size does not allow the investigation 
of the mixing induced CP-violation, a class of CP-violating effects which can reveal the 
presence of NP are the Triple Products ($TP$) correlations~\cite{phiphi_tp}. 
$TP$'s are defined as: $TP = {\vec p} \cdot ({\vec q_1} \times {\vec q_2})$ where ${\vec p}$ is 
a momentum, and ${\vec q_1}$ and ${\vec q_2}$ can be either the spins or momenta of the decay 
particles. Triple products are odd variables under time reversal (T), therefore they constitute 
potential signals of CP violation. The $TP$ asymmetry is defined as:
\begin{center}
$A_{TP} = \frac{\Gamma(TP>0) - \Gamma(TP<0)}{\Gamma(TP>0) + \Gamma(TP<0)}$, 
\end{center}
where $\Gamma$ is the decay rate of the process in question. Most of these $TP$ asymmetries are 
expected to be small in the SM, but can be enhanced in the presence of NP in the decay. In the untagged 
case the $TP$ asymmetries are proportional to the so-called "true" $TP$ asymmetry, that is a true CP 
violating effect. In what follows, for shortness, we refer to them as $TP$ only. 

In the $B_s \rightarrow \phi \phi$ decays, there are two Triple Products: $TP_2$ is proportional 
to ${\rm Im}(A_{||} A_{\perp})$ and $TP_1$ is related to ${\rm Im}(A_0 A_{\perp})$. $TP_2$ can be probed through 
the observable $u =$ cos$\varphi$ sin$\varphi$, where $\varphi$ is the angle between the two $\phi$ meson decay planes. 
The asymmetry on $u$, $A_u$, is proportional to the asymmetry 
of $TP_2$ and is defined as: $A_u = (N^+ - N^-) / (N^ + N^-)$, where $N^+ (N^-)$ are the number of 
events with $u>0$ ($u<0$).
In a similar way, we define an asymmetry $A_v$ for the variable $v = $sin$\varphi$ if cos$\theta_1$ cos$\theta_2>0$ 
and $v =$ sin$(-\varphi)$ if cos$\theta_1$ cos$\theta_2 \le 0$. The asymmetry $A_v$ which is proportional to the 
asymmetry of $TP_1$.
The $u$ and $v$ distributions are shown in Fig.~\ref{fig_bs_phi_phi_tp} for sideband-subtracted signal events.

\begin{figure}[tb]
\includegraphics[width=65mm]{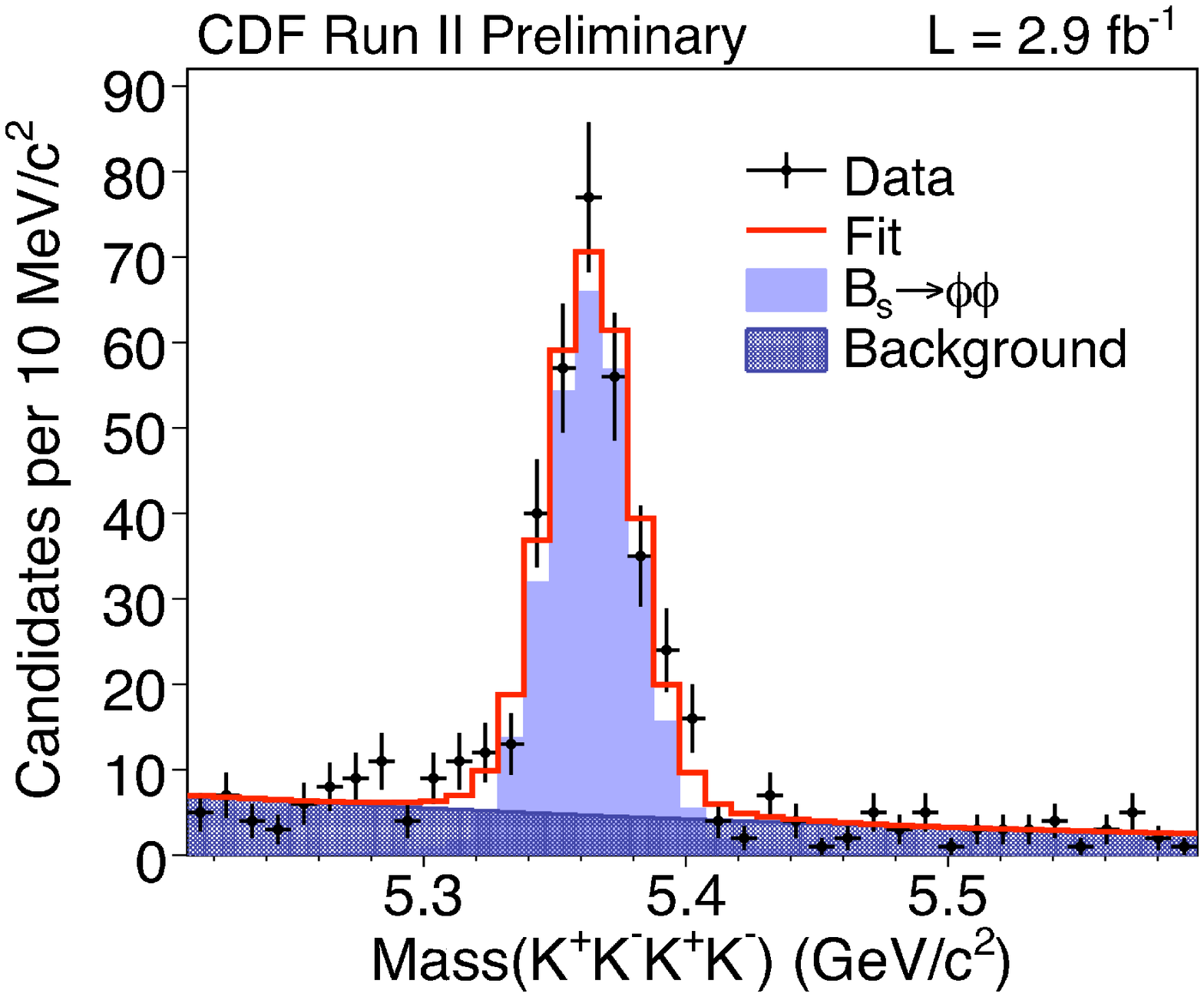}
\includegraphics[width=50mm]{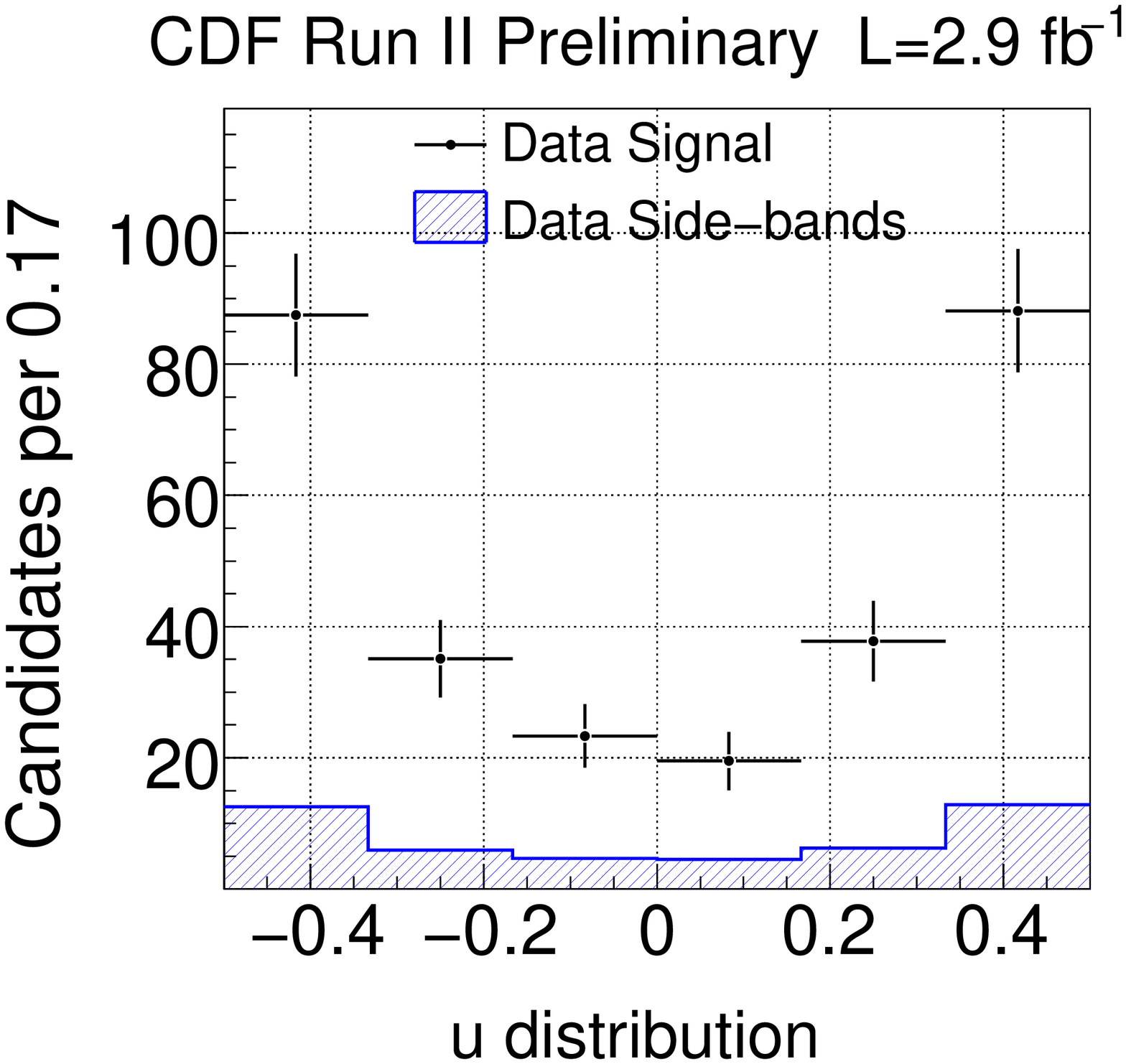}
\includegraphics[width=50mm]{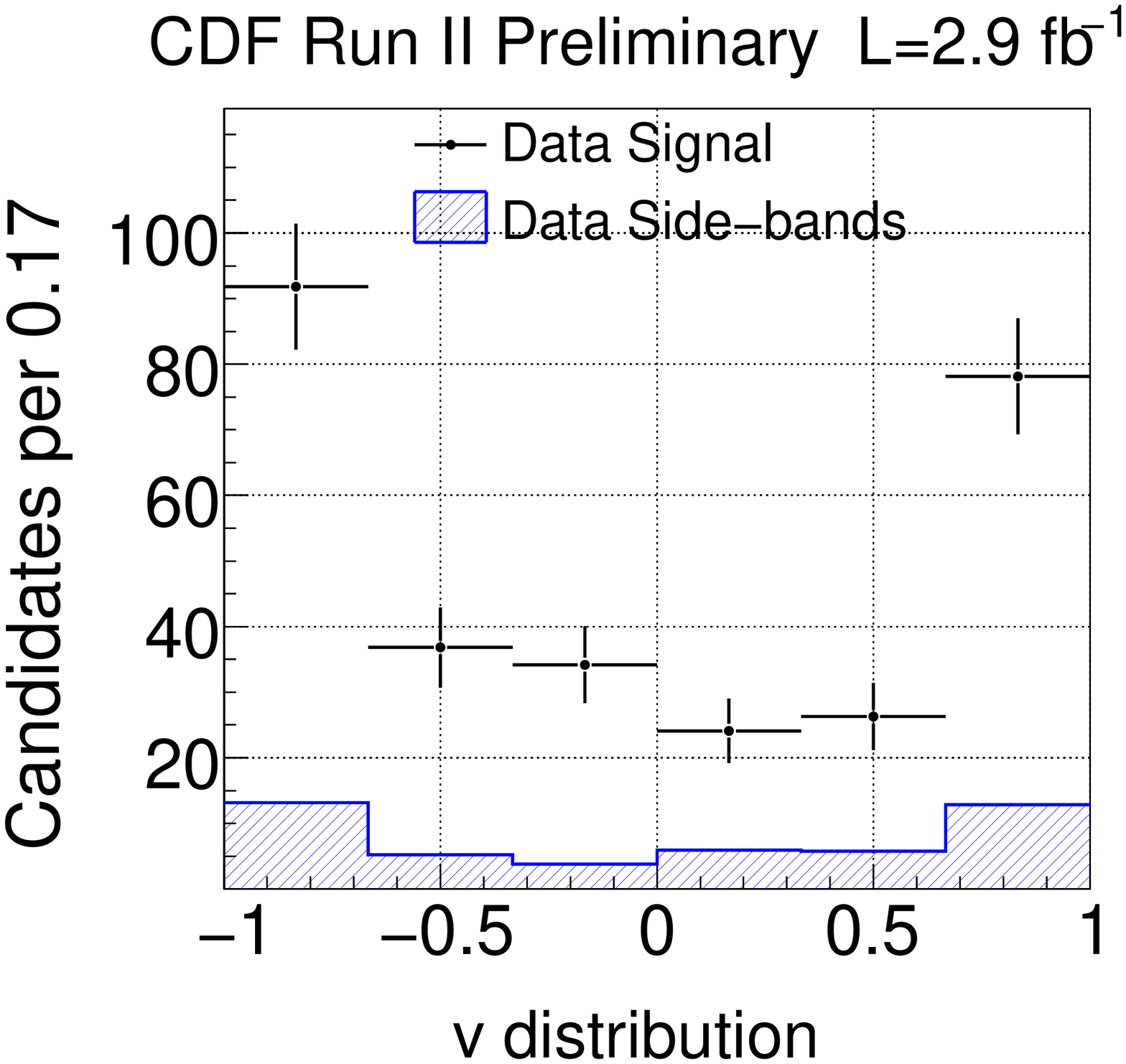}
\caption{ Invariant mass of $\phi(\rightarrow K^+K^-) \phi(\rightarrow K^+K^-)$ (left). 
	The $u$ (center) and $v$ (right) distributions in $B_s \rightarrow \phi \phi$ for 
	side-bands subtracted signal events.}
\label{fig_bs_phi_phi_tp}
\end{figure}

The measured asymmetries of the two T-odd helicity angles functions are:
\begin{center}
$A_u = -0.007 \pm 0.064 (\rm stat.) \pm 0.018 (\rm syst.)$, and \\
$A_v = -0.120 \pm 0.064 (\rm stat.) \pm 0.016 (\rm syst.)$.
\end{center}
The first asymmetry, $A_u$, is well consistent with zero within experimental uncertainties, while 
the second one, $A_v$, is 1.8 standard deviations from zero considering both statistical and systematic 
uncertainties. These asymmetries constrain the size of two T-violating true Triple Product asymmetries 
of the $B_s \rightarrow \phi \phi$ decay, expected null in the SM.

\section{CP Violation in $B \rightarrow DK$ Decays}

The branching fractions and CP asymmetries of $B^- \rightarrow D^0 K^-$ modes allow a theoretically-clean 
way of measuring the CKM angle $\gamma$, which is the least well-known CKM angle, with uncertainties of 
about 10-20 degrees. In particular, the ADS method"~\cite{ads1,ads2} makes use of modes where the $D^0$ decays 
in the doubly-Cabibbo-suppressed (DCS) mode: $D^0 \rightarrow K^+ \pi^-$. The large interference between the 
decays in which $B^-$ decays to $D^0 K^-$ through a color-favored $b \rightarrow c$ transition, followed by 
the DCS decay $D^0 \rightarrow K^+ \pi^-$, and the decay in which $B^-$ decays to $D^0 K^-$ through a 
color-suppressed $b \rightarrow u$ transition, followed by the Cabibbo-favored (CF) decay $D^0 \rightarrow K^+\pi^-$, 
can lead to measurable CP asymmetries, from which the $\gamma$ angle can be extracted.

The observables of the ADS method are:
\begin{center}
$R_{ADS}(K) = \frac{BR(B^- \rightarrow [K^+ \pi^-]_D K^-) + BR(B^+ \rightarrow [K^- \pi^+]_D K^+)}{BR(B^- \rightarrow [K^- \pi^+]_D K^-) + BR(B^+ \rightarrow [K^+ \pi^-]_D K^+)}$ \\

$A_{ADS}(K) = \frac{BR(B^- \rightarrow [K^+ \pi^-]_D K^-) - BR(B^+ \rightarrow [K^- \pi^+]_D K^+)}{BR(B^- \rightarrow [K^+ \pi^-]_D K^-) + BR(B^+ \rightarrow [K^- \pi^+]_D K^+)}$ \\

$R^{±}(K) = \frac{BR(B^{\pm} \rightarrow [K^{\mp} \pi^{\pm}]_D K^{\pm})}{BR(B^{\pm} \rightarrow [K^{\pm} \pi^{\mp}]_D K^{\pm})}$ \\
\end{center}
$R_{ADS}(K)$ and $A_{ADS}(K)$ are related to the $\gamma$ angle through these relations:
\begin{center}
$R_{ADS}(K) = r_D^2 + r_B^2 + r_Dr_B$ cos$\gamma$ cos$(\delta_B + \delta_D)$ and \\ 
$A_{ADS}(K) = 2 r_B r_D$ sin$\gamma$ sin$(\delta_B + \delta_D) / R_{ADS}(K)$
\end{center}
where $r_B = |A(b \rightarrow u) / A(b \rightarrow c)|$ and $\delta_B = {\rm arg}[A(b \rightarrow u) / A(b \rightarrow c)]$. 
$r_D$ and $\delta_D$ are the corresponding amplitude ratio and strong phase difference of the $D$ meson decay amplitudes.
As can be seen from the expressions above, $A_{ADS} ({\rm max}) = 2r_B r_D / (r_B^2 + r_D^2)$ is the maximum size of the asymmetry. 
For given values of $r_B(\pi)$ and $r_D$, sizeable asymmetries may be found also for $B^- \rightarrow D^0 \pi^-$ decays, 
so interesting observables are:
\begin{center}
$R_{ADS}(\pi) = \frac{BR(B^- \rightarrow [K^+ \pi^-]_D \pi^-) + BR(B^+ \rightarrow [K^- \pi^+]_D \pi^+)}{BR(B^- \rightarrow [K^- \pi^+]_D \pi^-) + BR(B^+ \rightarrow [K^+ \pi^-]_D \pi^+)}$ \\
$A_{ADS}(\pi) = \frac{BR(B^- \rightarrow [K^+ \pi^-]_D \pi^-) - BR(B^+ \rightarrow [K^- \pi^+]_D \pi^+)}{BR(B^- \rightarrow [K^+ \pi^-]_D \pi^-) + BR(B^+ \rightarrow [K^- \pi^+]_D \pi+)} $ \\
$R^{\pm}(\pi) = \frac{BR(B^{\pm} \rightarrow [K^{\mp} \pi^{\pm}]_D \pi^{\pm})}{BR(B^{\pm} \rightarrow [K^{\pm} \pi^{\mp}]_D \pi^{\pm})}$
\end{center}

The CDF experiment presents an ADS analysis~\cite{cdf_b_dk} on a data sample corresponding to 7~fb${^-1}$ 
of integrated luminosity. 
An extended maximum likelihood fit that combines mass and particle identification information is used 
to separate statistically the $B^- \rightarrow D K^-$ contributions from 
the $B^- \rightarrow D\pi^-$ signals and from the combinatorial and physics backgrounds.
The $B^- \rightarrow D \pi^-$ signal is reconstructed with a statistical significance of 3.6~Gaussian~sigma. 
The suppressed signals $B^- \rightarrow D K^-$ are reconstructed with a significance of 3.2~sigma, including 
systematics. The plots in Fig.~\ref{BDK} show the $B$ invariant mass distribution for positive and negative charges 
of the suppressed sample. 

\begin{figure}[tb]
\includegraphics[width=55mm]{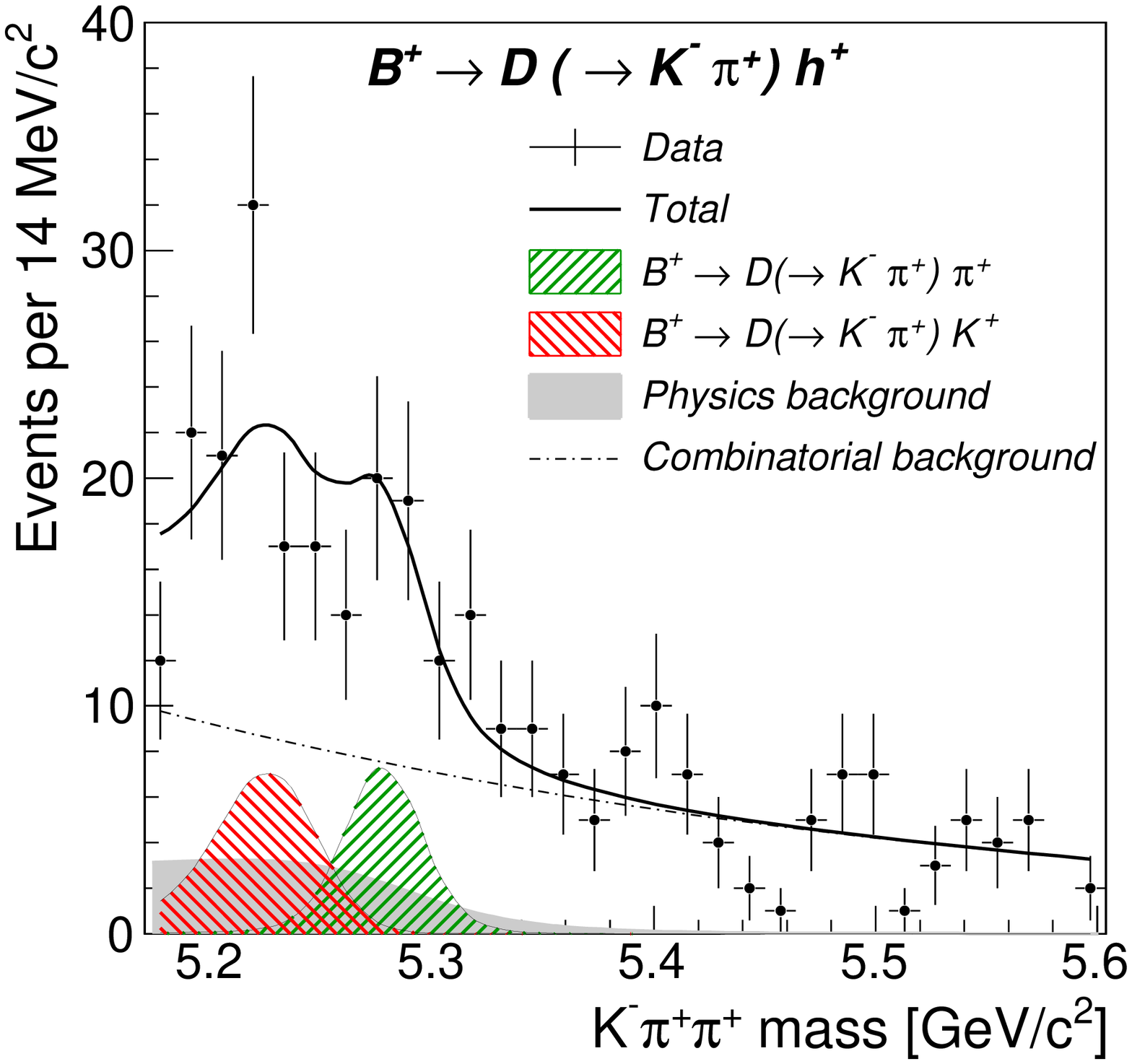}
\includegraphics[width=55mm]{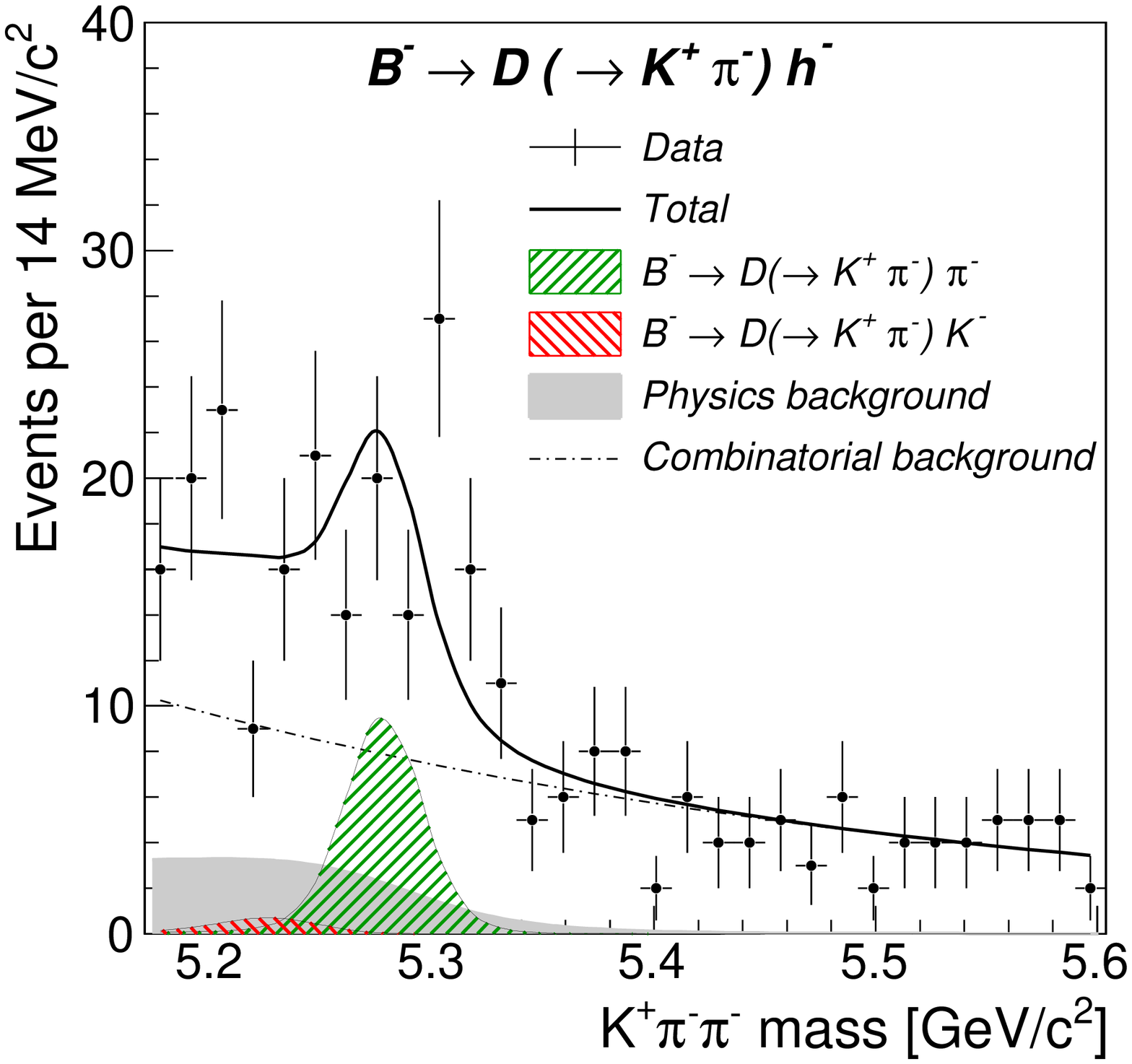}
\includegraphics[width=55mm]{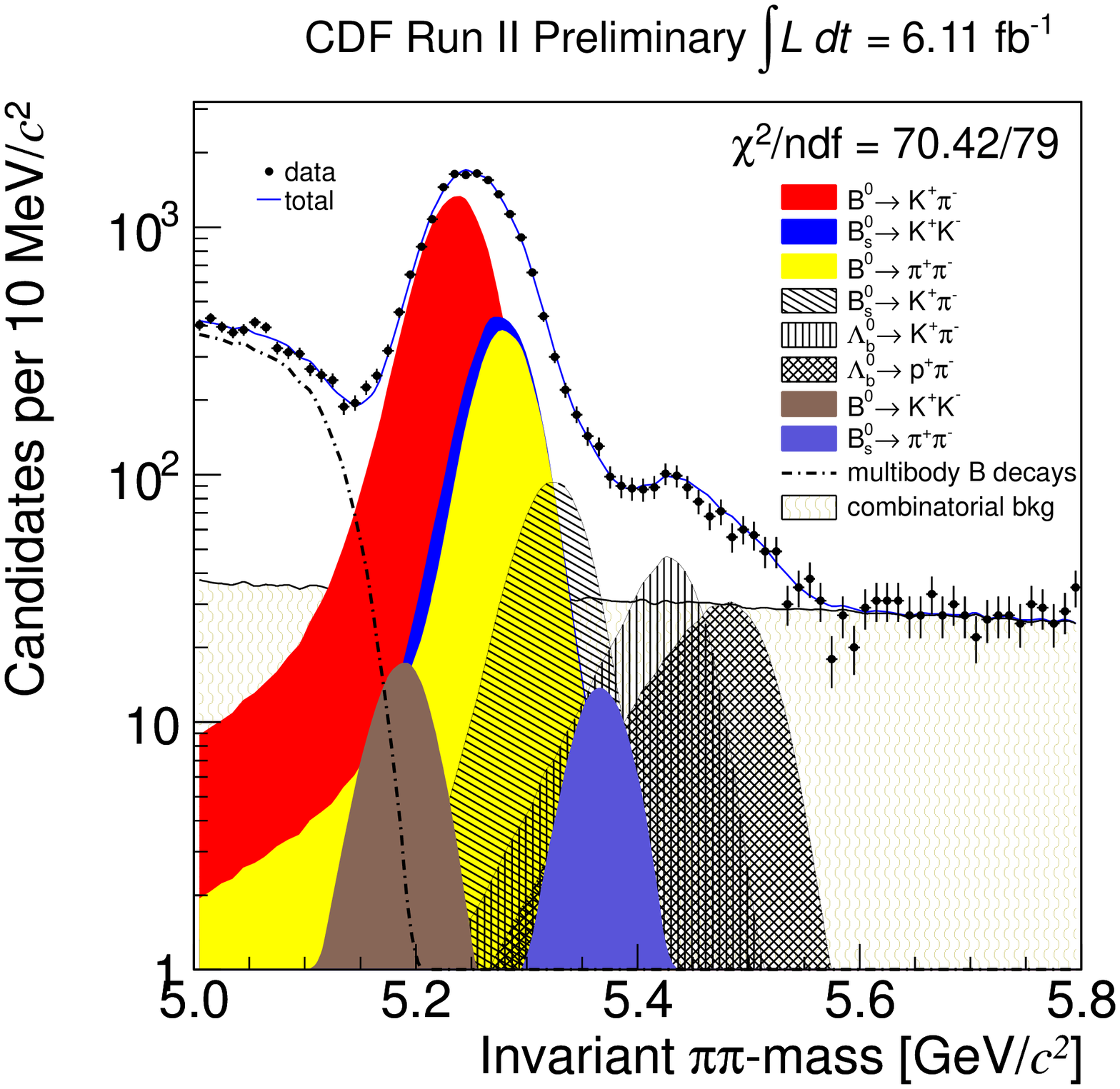}
\caption{ Invariant mass distribution of $B^- \rightarrow D_{\rm suppressed} h^-$ for positive (left)   
		and negative charges (center); the $h^-$ hadron is either a $K^-$ or a $\pi^-$.  
		Invariant mass of $\pi^+ \pi^-$ from $B \rightarrow h^+ h^-$ candidates, 
	     	where the $h$ hadron is either a pion or a kaon (right).} 
\label{BDK}
\end{figure}

The ratios of the suppressed to favored branching fractions are measured as:
\begin{center}
$R_{ADS}(K) = [22.0 \pm  8.6(\rm stat.) \pm 2.6(\rm syst.)] \times 10^{-3}$, \\ 
$R^+(K)     = [42.6 \pm 13.7(\rm stat.) \pm 2.8(\rm syst.)] \times 10^{-3}$, \\ 
$R^-(K)     = [ 3.8 \pm 10.3(\rm stat.) \pm 2.7(\rm syst.)] \times 10^{-3}$, \\ 

$R_{ADS}(\pi) = [ 2.8 \pm  0.7(\rm stat.) \pm 0.4(\rm syst.)] \times 10^{-3}$, \\ 
$R^+(\pi)     = [ 2.4 \pm  1.0(\rm stat.) \pm 0.4(\rm syst.)] \times 10^{-3}$, \\ 
$R^-(\pi)     = [ 3.1 \pm  1.1(\rm stat.) \pm 0.4(\rm syst.)] \times 10^{-3}$,

\end{center}
as well as the direct CP-violating asymmetry 
\begin{center}
$A_{ADS}(K)   = -0.82\pm 0.44(\rm stat.) \pm 0.09(\rm syst.)$, \\ 
$A_{ADS}(\pi) =  0.13\pm 0.25(\rm stat.) \pm 0.02(\rm syst.)$.
\end{center}

The results are in agreement and competitive with $B$-factories~\cite{ads_bfac} and 
with the LHCb experiment~\cite{ads_lhcb}. 

\section{ Two Body Charmless B Decays }

The decay modes of $B$-mesons into pairs of charmless pseudo-scalar mesons are effective 
probes of the quark-mixing (CKM) matrix and are sensitive to potential new physics effects. 
Their branching fractions and CP asymmetries can be predicted with good accuracy and compared 
to the rich experimental data available for $B_u$ and $B_d$ mesons produced in large quantities 
in $\Upsilon(4S)$ decays~\cite{bhh1}. Measurements of similar modes predicted, 
for the $B_s$ meson are important to supplement our understanding of $B$-meson decays. 
The measurement of observables from both strange and non-strange $B$-mesons allows a cancellation 
of hadronic uncertainties, thus enhancing the precision of the extraction of physics parameters 
from experimental data~\cite{bhh2,bhh3,bhh4,bhh5}.
A combination of $B^0 \rightarrow \pi^+ \pi^-$ and $B_s \rightarrow K^+K^-$ observables has been proposed 
as a way to directly determine the phase of the $V_{ub}$ element of the CKM matrix (angle $\gamma$), 
or alternatively as a test of our understanding of the dynamics of $B$ hadron decays, when compared 
with other determinations of $\gamma$~\cite{bhh6}. The $B_s \rightarrow K^-\pi^+$ mode can also be used in 
measuring $\gamma$~\cite{bhh3}, and its CP asymmetry is a powerful model-independent test~\cite{bhh7} 
of the source of the direct CP asymmetry recently observed in the $B^0 \rightarrow K^+ \pi^-$ mode~\cite{bhh8}. 
The $B_s \rightarrow \pi^+ \pi^-$ mode proceeds only through annihilation diagrams, which are currently 
poorly known and a source of significant uncertainty in many theoretical calculations~\cite{bhh9}. 
Its features are similar to the $B^0 \rightarrow K^+ K^-$ mode, but it has a larger 
predicted branching fraction~\cite{bhh10}; a measurement of both modes would allow a determination of the 
strength of penguin-annihilation~\cite{bhh4}. 

Channels previously investigated by the CDF experiment are 
$B_s \rightarrow K^+K^-$~\cite{cdf_bhh1},  
$B_s \rightarrow K^- \pi^+$,
$\Lambda^0_b \rightarrow p \pi^-$ and  
$\Lambda^0_b \rightarrow p K^-$~\cite{cdf_bhh2}, 
and the corresponding asymmetries 
$A_{CP}(B_s \rightarrow K^- \pi^+)$, 
$A_{CP}(\Lambda^0_b \rightarrow p \pi^-)$ and  
$A_{CP}(\Lambda^0_b \rightarrow p K^-)$~\cite{cdf_bhh3}.

Recently, the CDF experiment has established~\cite{bs_pipi} the 
first evidence for $B_s \rightarrow \pi^+ \pi^-$ decays and 
has set bounds on the branching fraction of the $B^0 \rightarrow K^+K^-$ decay mode. 
Fig.~\ref{BDK} shows the invariant mass of $\pi^+ \pi^-$ from 
$B \rightarrow h^+ h^-$ candidates. The $B_s \rightarrow \pi^+ \pi^-$ and 
$B^0 \rightarrow K^+K^-$ are $94 \pm 28 (\rm stat.) \pm 11 (\rm syst.)$ 
and $120 \pm 49 (\rm stat.) \pm 42 (\rm syst.)$, respectively. The branching fractions are measured as:
\begin{center}
$BR(B_s \rightarrow \pi^+ \pi^-) = (0.57 \pm 0.15 (\rm stat.) \pm 0.10 (\rm syst.)) \times 10^{-6}$, \\
$BR(B^0 \rightarrow   K^+   K^-) = (0.23 \pm 0.10 (\rm stat.) \pm 0.10 (\rm syst.)) \times 10^{-6}$, \\
	$BR(B^0 \rightarrow   K^+   K^-) \in [0.05, 0.46] \times 10^{-6}$ at $90\%$~C.L. 
\end{center}

\section {Observation of $\Xi^0_b$}

The Tevatron experiments, D0 and CDF, have had major contribution to $b$-baryon 
spectroscopy, with the observations of the $\Xi^{-}_{b} (dsb)$~\cite{bb1}, 
$\Sigma_b^{*} (uub, ddb)$~\cite{bb2} and $\Omega_b (ssb)$~\cite{bb2} baryons. 

We report the observation by the CDF experiment of an additional heavy baryon, 
$\Xi^0_b (usb)$~\cite{cdf_Xi_b} and 
the measurement of its mass. The measurement uses 4.2~fb$^{-1}$ of integrated 
luminosity. The $\Xi^0_b$ baryon is observed through its decay 
\begin{center}
$\Xi_b^0 \rightarrow \Xi_c^+ \pi^-$, where \\
$\Xi_c^+ \rightarrow \Xi^- \pi^+ \pi^+$, $\Xi^- \rightarrow \Lambda \pi^-$ 
and $\Lambda \rightarrow p \pi^-$. 
\end{center}
In addition, the $\Xi_b^-$ baryon is observed through 
the similar decay chain
\begin{center}
$\Xi_b^- \rightarrow \Xi_c^0 \pi^-$, where \\
$\Xi_c^0 \rightarrow \Xi^- \pi^+$, $\Xi^- \rightarrow \Lambda \pi^-$ and $\Lambda \rightarrow p \pi^-$.
\end{center}
The $\Xi_b^0$ and $\Xi_b^-$ candidate mass distributions are shown in Fig.~\ref{fig:xi}. 
There are $25.3^{+5.6}_{-5.4}$~$\Xi_b^0$ candidates and $25.8^{+5.5}_{-5.2}$~$\Xi_b^-$ candidates 
with measured masses of $5787.8 \pm 5.0 (\rm stat.) \pm 1.3 (\rm syst.)$~MeV/c$^2$ and
$5796.7 \pm 5.1 (\rm stat.) \pm 1.4 (\rm syst.)$~MeV/c$^2$, respectively.
The $\Xi_b^0$ signal significance is greater than 6$\sigma$. Neither of these 
decay channels has been reported previously and the reconstruction of $\Xi_b^0$ is the 
first observation of this baryon in any channel. 

\begin{figure}[tb]
\includegraphics[width=65mm]{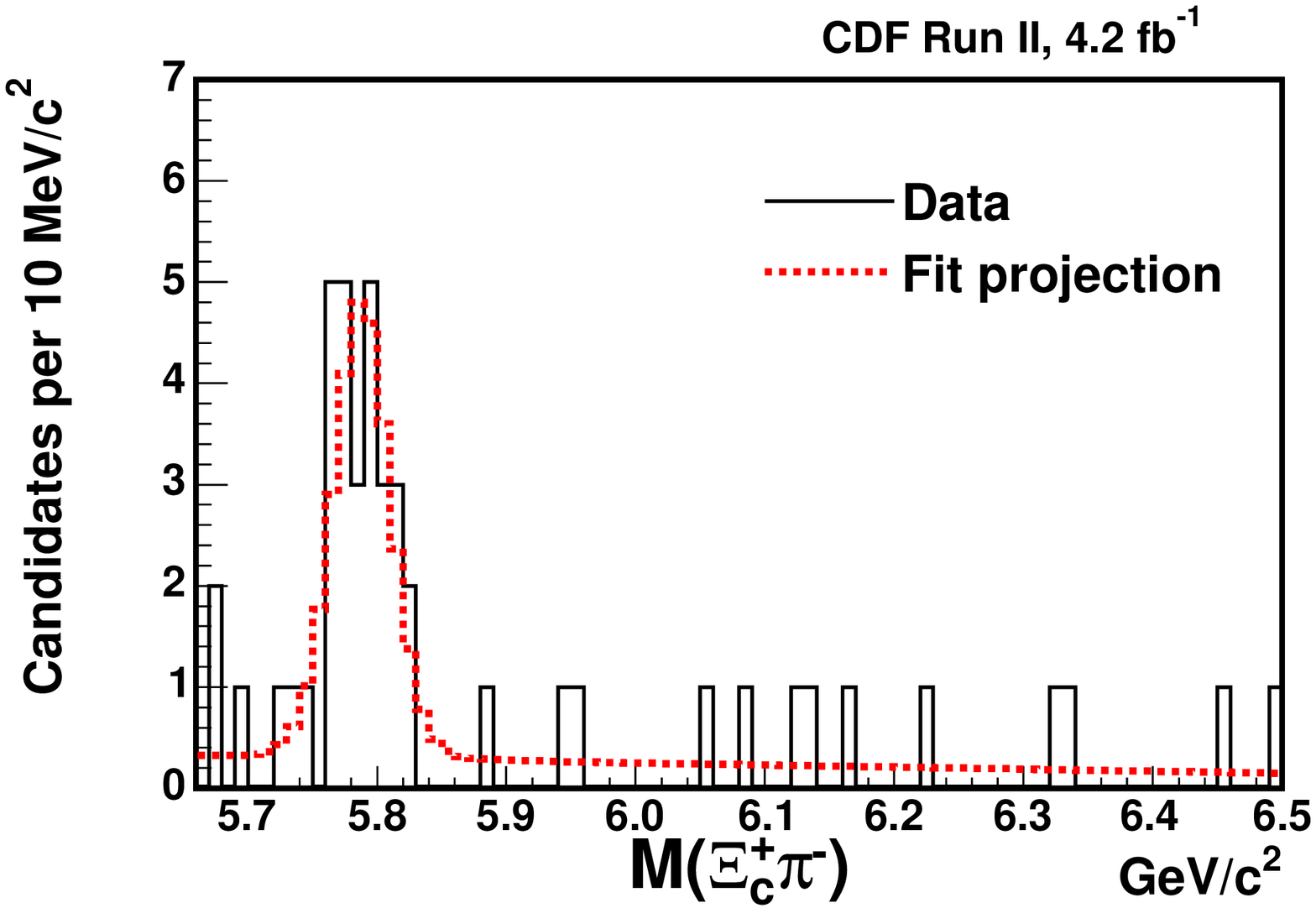}
\hskip0.5in
\includegraphics[width=65mm]{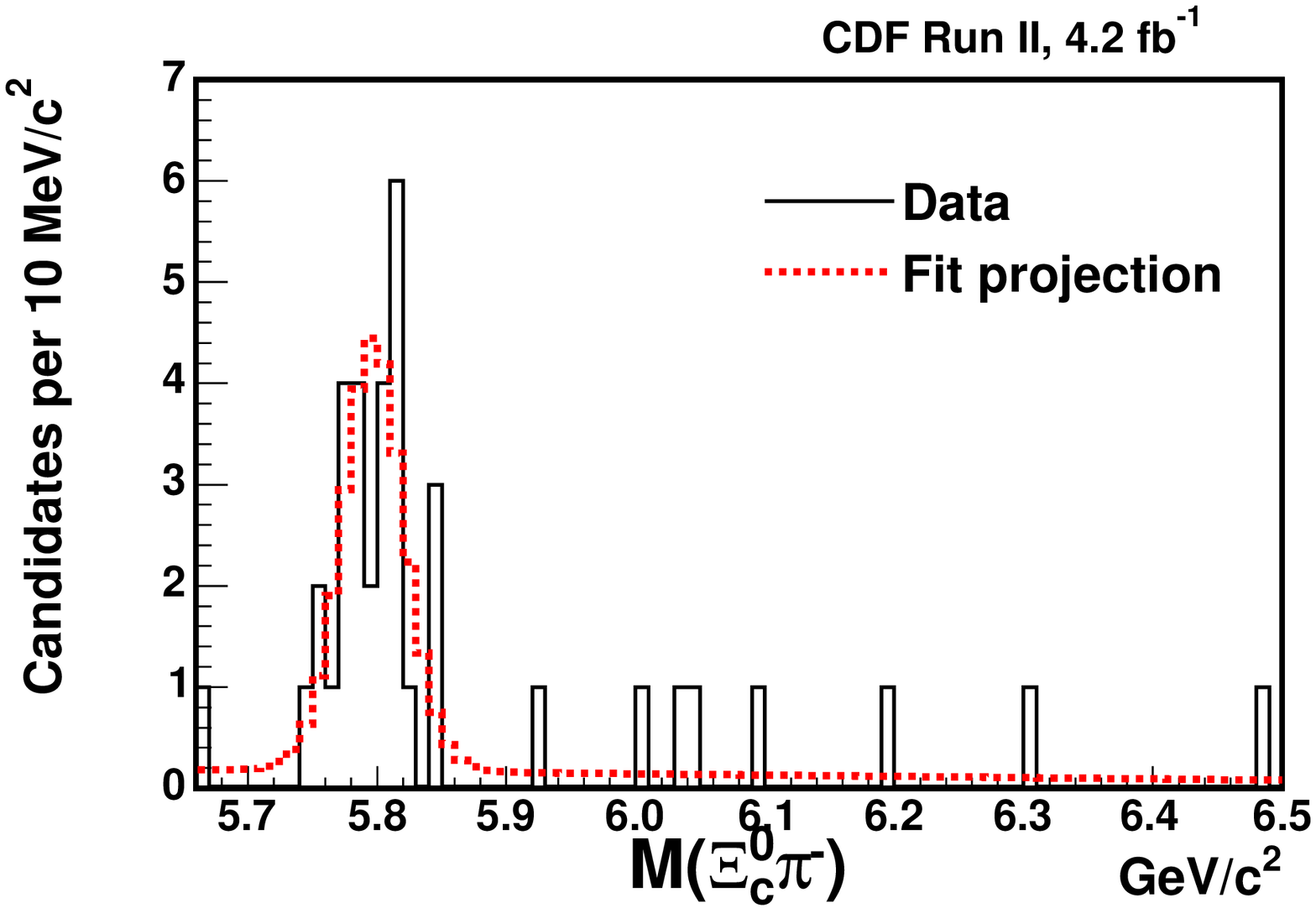}
\caption{ Invariant mass distribution of $\Xi_c^+ \pi^-$ (left) and $\Xi_c^0 \pi^-$ (right) with overlaid fit projection. }
\label{fig:xi}
\end{figure}

\section {Conclusions}
The D0 and CDF experiments are continuing to produce a rich and exciting 
program in heavy flavor physics: 
interesting effects in same-sign di-muon asymmetry and $B_s \rightarrow \mu^+ \mu^-$ decays, 
as well as the best measurements of CP-violating phase, $\beta_s/\phi_s$.  
Many interesting results will benefit from increasing the data samples.
It is anticipated that each of the two Tevatron experiments will accumulate 
approximately 10~fb$^{-1}$ of integrated luminosity by the end of the Tevatron run.

\end{document}